\documentclass[twocolumn,english,aps,pra,tightenlines,floatfix,showpacs]{revtex4-1}
\usepackage{amsmath}
\usepackage{amssymb}
\usepackage{graphicx}
\usepackage{esint}

\makeatletter
\@ifundefined{textcolor}{}
{%
 \definecolor{BLACK}{gray}{0}
 \definecolor{WHITE}{gray}{1}
 \definecolor{RED}{rgb}{1,0,0}
 \definecolor{GREEN}{rgb}{0,1,0}
 \definecolor{BLUE}{rgb}{0,0,1}
 \definecolor{CYAN}{cmyk}{1,0,0,0}
 \definecolor{MAGENTA}{cmyk}{0,1,0,0}
 \definecolor{YELLOW}{cmyk}{0,0,1,0}
}

\renewcommand{\[}{\begin{equation}}
\renewcommand{\]}{\end{equation}}

\usepackage{times}

\makeatother

\usepackage{babel}
\begin{document}
\global\long\def\avg#1{\langle#1\rangle}

\global\long\def\p{\prime}

\global\long\def\ket#1{|#1\rangle}

\global\long\def\bra#1{\langle#1|}

\global\long\def\proj#1#2{|#1\rangle\langle#2|}

\global\long\def\inner#1#2{\langle#1|#2\rangle}

\global\long\def\tr{\mathrm{tr}}

\global\long\def\dg{\dagger}

\global\long\def\im{\imath}

\global\long\def\pd#1#2{\frac{\partial#1}{\partial#2}}

\global\long\def\spd#1#2{\frac{\partial^{2}#1}{\partial#2^{2}}}

\global\long\def\der#1#2{\frac{d#1}{d#2}}

\renewcommand{\thefootnote}{\fnsymbol{footnote}}
\setcounter{footnote}{1}

\title{Landauer, Kubo, and microcanonical approaches to quantum transport and noise: A comparison and implications for cold-atom dynamics}

\author{Chih-Chun Chien$^{1}$, Massimiliano Di Ventra$^{2}$, and Michael
Zwolak$^{3}$}

\affiliation{$^{1}$School of Natural Sciences, University of California, Merced, CA 95343, USA \\
 $^{2}$Department of Physics, University of California, San Diego,
CA 92093, USA \\
 $^{3}$Department of Physics, Oregon State University, Corvallis,
OR 97331, USA }

\date{\today}
\begin{abstract}
We compare the Landauer, Kubo, and microcanonical [J. Phys. Cond. Matter {\bf 16}, 8025 (2004)] approaches to quantum transport  for the average current, the entanglement entropy and the semiclassical full-counting statistics (FCS). Our
focus is on the applicability of these approaches to isolated quantum systems such as ultra-cold atoms in
engineered optical potentials. For two lattices connected by a junction,
we find that the current and particle number fluctuations from the
microcanonical approach compare well with the values predicted by the Landauer
formalism and FCS assuming a binomial distribution. However, we
demonstrate that well-defined reservoirs (i.e., particles in Fermi-Dirac distributions)
are not present for a substantial duration of the quasi-steady
state. Thus, on the one hand, the Landauer assumption of reservoirs and/or inelastic effects is not necessary for establishing a quasi- steady state. Maintaining such a state indefinitely requires an infinite system, and in this limit well-defined Fermi-Dirac distributions can occur. On the other hand, as we show, the existence of a finite speed of particle propagation preserves the quasi-steady state irrespective of the existence of well-defined reservoirs. This indicates that global observables in finite systems may be substantially different from those predicted by an uncritical application of the Landauer formalism, with its underlying thermodynamic limit. Therefore, the microcanonical formalism
which is designed for closed, finite-size quantum systems seems 
more suitable for studying particle dynamics in ultra-cold atoms.
Our results highlight both the connection and differences with more traditional
approaches to calculating transport properties in condensed matter
systems, and will help guide the way to their simulations in cold-atom
systems.
\end{abstract}


\maketitle

\section{Introduction}
Experimental investigations of transport phenomena in ultra-cold atoms
confined in engineered optical potentials offer a test bed for transport theories at the nanoscale. Several phenomena, such as the sloshing
motion of an atomic cloud in optical lattices \cite{Ott04}, directed
transport using a quantum ratchet \cite{Salger09}, relaxation of noninteracting and interacting fermions in optical lattices \cite{Blochtransport}, and others have been demonstrated.
Their applications in atomtronics \cite{Atomtronics1}, which aims
at simulating electronics by using controllable atomic systems, are
promising \cite{ring1,Esslinger12,Hill13,Zozulya,Eckel14}.
It is thus important to develop proper theoretical and
computational methods to direct future progress in this field.

Due to the quantum nature of atoms, finite particle numbers, and small
sizes of these systems, the applicability of semi-classical approaches,
such as the Boltzmann equation, become questionable. The Landauer formalism \cite{Landauer,DiVentrabook},
which has been widely implemented in mesoscopic physics, is naturally
appealing for studying transport phenomena in ultra-cold atoms.
Those approaches and their generalizations have been applied to study various problems in
cold atoms \cite{Bruun05,SchaferRoPP,Brantut,Bruderer12,Gutman12,Simpson14}.
In addition to steady-state properties, one may want to study fluctuation
effects and correlations using full-counting statistics (FCS) \cite{KlichLevitov}. An examination of the underlying
assumptions of those well-known formalisms, however, raise questions on their applicability to ultra-cold
atoms.

The Landauer formalism, which is designed for open systems, assumes
the existence of two reservoirs that supply particles to be transmitted
through a junction region. Since the particle number and energy (when no external time-dependent fields are present) in
ultra-cold atomic experiments are (to a very good approximation) conserved,
the concept of a reservoir does not necessarily hold. FCS generally assumes the transmitted
particles behave like billiards with a well-defined tunneling probability
distribution. Whether such an assumption holds true in finite, closed
systems will determine whether the formalism can be applied to cold
atom experiments as well.

An alternative approach for studying transport in quantum systems
is within the {\it microcanonical formalism} (MCF) \cite{micro,Bushong05,DiVentrabook}.
This formalism is based on using closed quantum systems driven out
of equilibrium by a change of parameters (e.g., an external bias or a density imbalance) to calculate transport
properties. The conservation of particle number and energy are naturally
built into this formalism, and there is no need to introduce reservoirs and one can fully
preserve the wave nature of the particles.
This formalism has also been integrated with density-functional theory for investigating quantum transport
through atomic or molecular junctions \cite{Gross05,VanVoorhis06}.
The microcanonical formalism is particularly suitable
for ultra-cold atoms, which are accurately modeled as isolated quantum systems.
In this respect, the formalism has already been developed
to study transport phenomena in these systems \cite{MCFshort,MCF_TD,PhaseTran13,int_induced,GW14,Peotta2014}.

\begin{figure}
\begin{centering}
\includegraphics[width=2.3in]{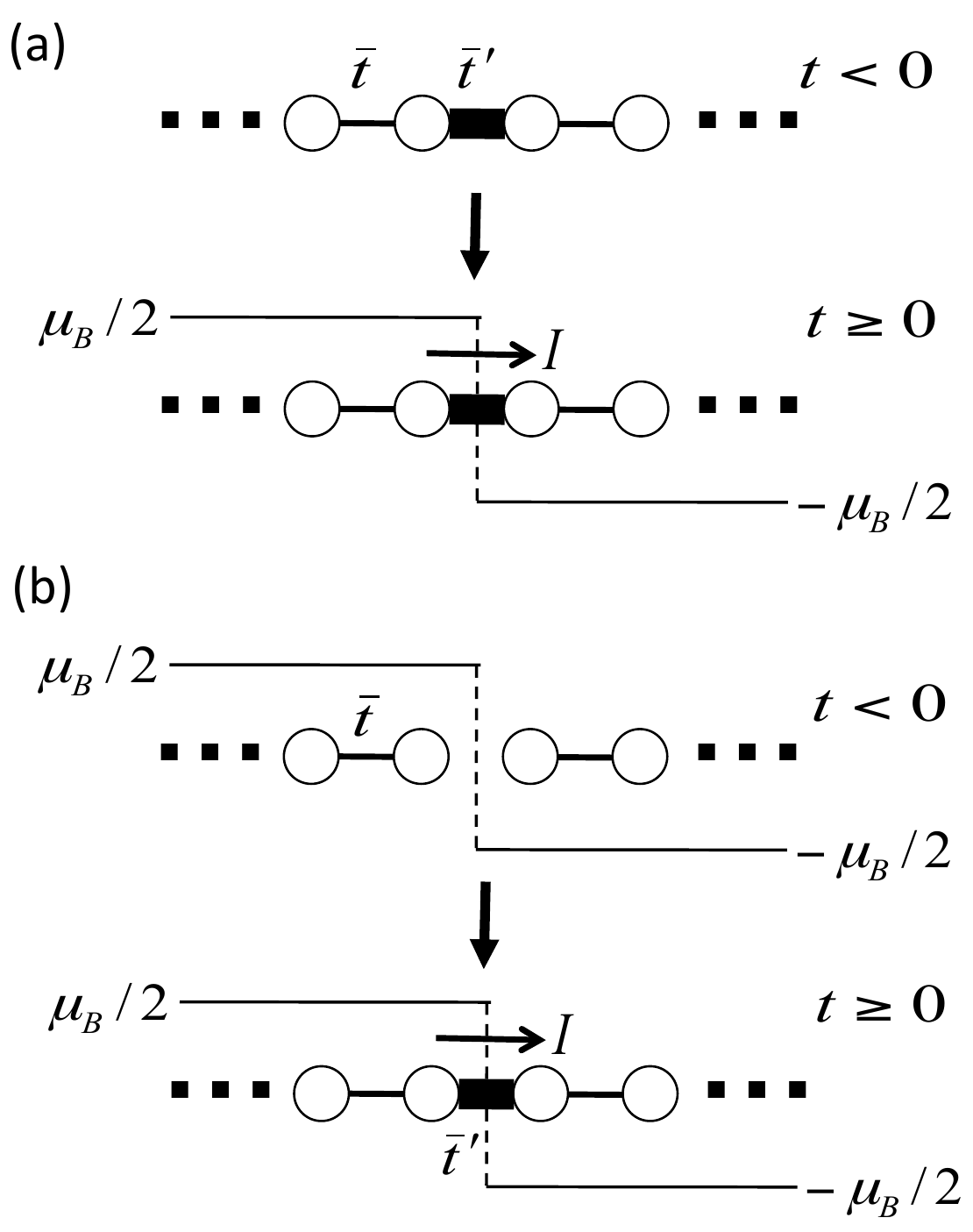}
\par\end{centering}
\caption{Schematic of a one dimensional lattice and transport induced by (a) application of a bias at $t=0$, where a step-function
bias is applied to the system and a current flows through the middle
link or (b) connecting a link between the two initially disconnected parts at $t=0$.
 \label{fig:Schematic}}
\end{figure}

The goal of this paper is to compare the microcanonical approach
to the Landauer formalism and determine which assumptions lead to the same observables, such as the average
current and FCS. The MCF is generically applicable to closed quantum systems, and here we use transport
of ultra-cold non-interacting fermions in one-dimensional (1D) optical
lattices as a particular example.
A possible setup is shown in Figure \ref{fig:Schematic}.
Unlike electronic systems where the Coulomb interactions cannot be
really switched off, and therefore for which this comparison would be more academic, cold atoms experiments allow for
a relatively easy tuning of interactions among particles down to the non-interacting limit. The microcanonical formalism can, of course, be applied to systems with Coulomb interactions. However, here we focus only on its applications to noninteracting cold-atom systems. While electrons are naturally confined in solid-state systems,
 a background harmonic trapping potential is often implemented in addition to the optical lattice for confining atoms. However, recent advance in trapping atoms in ring-shape geometries \cite{ring1,ring2} or a uniform potential \cite{Boxpotential} makes it possible to consider homogeneous cold-atom systems. Moreover, a weak background harmonic potential does not change the qualitative conclusions from the MCF, as illustrated in Ref.~\cite{MCFshort}. Therefore we focus here on the dynamics of cold atoms in optical lattices without a background harmonic trapping potential.

We find that the steady-state current and particle number fluctuations
from the microcanonical formalism approach the values of the average current and FCS predicted from the
Landauer formalism already at moderate system sizes. However,
we also find -- for finite times -- that one of the assumptions of the Landauer formalism is unnecessary:
The particle distributions in the two lattices supplying/absorbing
particles do not need to be populated according to Fermi-Dirac distributions. In fact, their occupation deviates
from the equilibrium distribution during the whole duration of a quasi-steady state. Furthermore, the results from the
microcanonical formalism agree with the predictions from the FCS semi-classical
formula by assuming a binomial distribution of the transmitted particles, ruling out alternative semi-classical descriptions. To connect the
different approaches, we also develop a Kubo formalism based on the microcanonical
picture of transport, which we use to calculate explicit expressions
for transport in closed systems. This gives us an analytical method to investigate
dynamical transport phenomena in nanoscale and ultracold atomic systems.

In addition to the average current and FCS, we also investigate the dynamical evolution
of the entanglement entropy, which quantifies the correlations between
two connected systems. The entanglement entropy is of broad interest in many fields,
ranging from black hole physics \cite{holography_book} to quantum
information science \cite{QCbook}. This quantity can be easily evaluated
using the microcanonical formalism. A semi-classical formula based
on FCS of two noninteracting fermionic systems connected by a junction
has been derived in Ref.~\cite{KlichLevitov} and generalized to many-body systems \cite{KlichPRB11,KlichPRB12}.
Ref.~\cite{KlichLevitov} predicts a linear
growth of the entanglement entropy as time increases. Again, we find that
the results from the microcanonical formalism match the prediction
from the semi-classical formula by assuming a binomial distribution
of the transmitted particles. Assuming an alternative distribution
results in predictions that are readily distinguishable.

This paper is organized as follows: Section~\ref{sec:Landauer} reviews the
Landauer formalism and its assumptions. Section~\ref{sec:MCF} introduces the
microcanonical formalism and its applications. The spatially resolved current from the MCF
is discussed in Section~\ref{sec:MCFspatial}. Section~\ref{sec:FCS} reviews the FCS. Section~\ref{sec:memory} shows the absence of memory effects in transport of
noninteracting fermions. Section~\ref{sec:comparison} compares the results from the MCF and
Landauer formalism. Importantly, the deviation from the equilibrium Fermi-Dirac distribution is clearly demonstrated. Section~\ref{sec:lightcone} shows the light-cone structure of wave propagation monitored
by the MCF. Section~\ref{sec:Kubo} reviews the Kubo
formalism and how it helps connect the two approaches. Finally, Section~\ref{sec:conclusion} concludes our study with suggestions of future work.

\section{Landauer formalism}\label{sec:Landauer}
By assuming the existence of a steady-state current between two reservoirs
bridged by a central link, the current can be estimated from the Landauer
formula with the help of, e.g., Green's functions \cite{Landauer,DiVentrabook}. For a detailed description
of the physical assumptions behind this formalism we refer the reader to Ref. \cite{DiVentrabook}.
Here, we mention only the assumptions that will be relevant for our comparison with the microcanonical formalism: (1) A steady-state current
is \textit{assumed} to exist. Whether a steady-state current always emerges from a given nonequilibrium condition
is not at all obvious \cite{DiVentrabook}. (2) Two macroscopic reservoirs -- holding noninteracting fermions populated according to Fermi-Dirac distributions -- are also assumed.
The separation of the
system into reservoirs and a region of interest is not always easy to determine for an actual physical structure. (3) The transport at the junction does not provide any feedback to the
reservoirs.

While one can construct
configurations where a steady-state current does not exist \cite{noQSSC_note}, in the
case where two 1-D chains are connected by a central junction (as considered
in this paper), there is always a steady-state current, as will be verified in
the microcanonical formalism (see Section~\ref{sec:MCFspatial}). Therefore, we do not focus here on assumption (1), but
rather on (2) and (3). As will be shown in Sec.~\ref{sec:comparison}, the distributions on both sides deviates from the Fermi-Dirac distribution when the system maintains a steady state so assumption (2) is not necessary for observing a steady-state current. Moreover, Section~\ref{sec:lightcone} will show that density changes can propagate into regimes away from the junction so there can be feedback and assumption (3) is also not necessary.

\begin{figure}
\includegraphics[clip,width=3.4in]{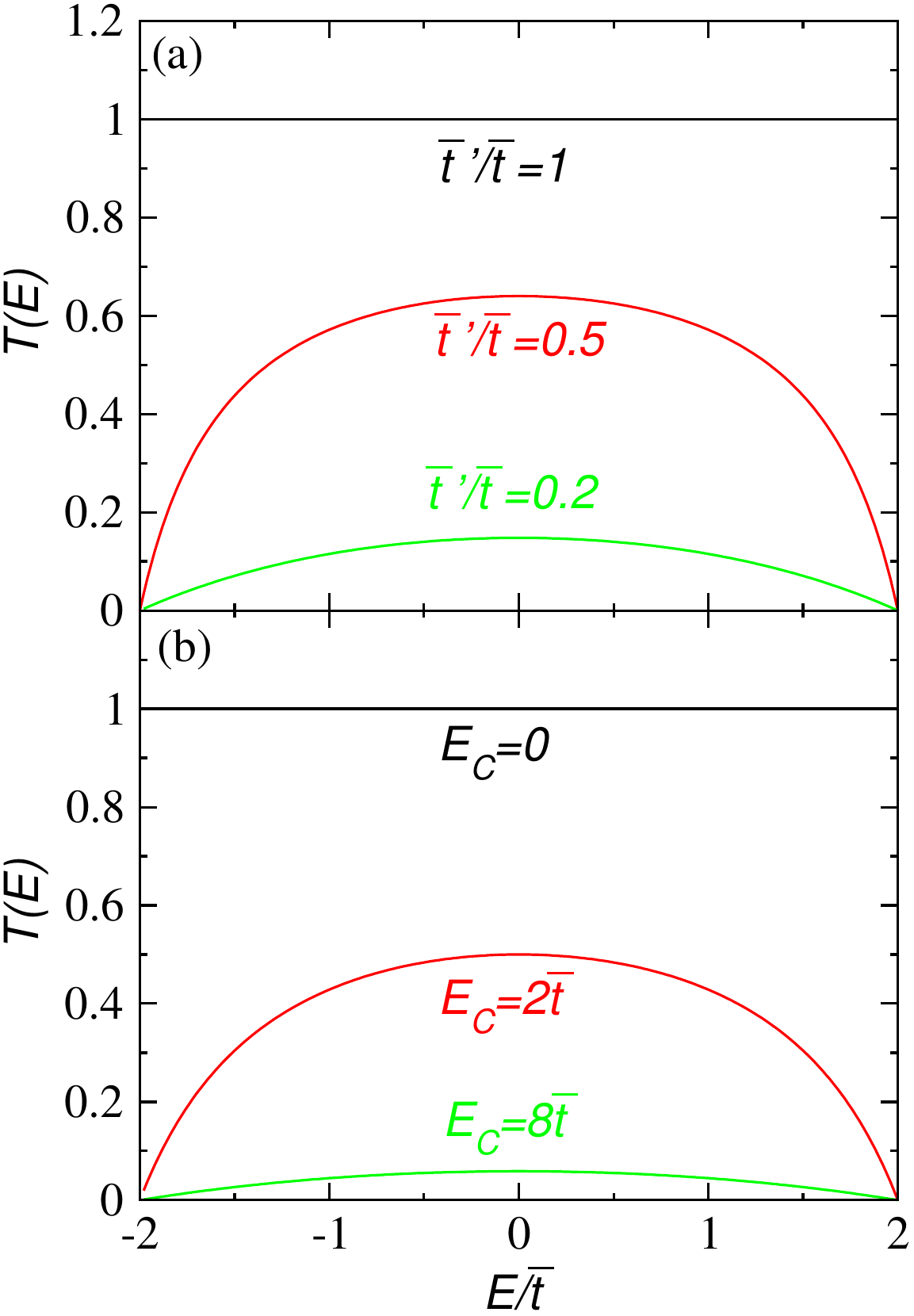} \caption{(Color online) The transmission coefficients $T(E)$ at $\mu_B=0$ for (a) the weak-link
case ($\bar{t}^{\prime}/\bar{t}=1,0.5,0.2$ from top to bottom)
and (b) the central-site case ($E_{C}/\bar{t}=0,2,8$ from top to
bottom).}
\label{fig:TE_2cases}
\end{figure}

On the other hand, in this section we calculate the current using the Landauer formalism for two configurations of a junction between two
1D lattices. One can insert a link with a tunable hopping coefficient
$\bar{t}^{\prime}$ in the middle of a chain, which we call the
{\it weak-link case}, or insert a central site with tunable on-site energy
$E_{C}$, which we call the {\it central-site case}.
In cold-atom experiments it has been shown that one can suppress the transmission of atoms by introducing an optical barrier \cite{ring1} or by introducing a constriction in the trapping potential \cite{Esslinger12}. Therefore the tunneling coefficient and onsite energy may be tuned simultaneously. Here we separate the effects of tuning the two parameters and one will see that there is no observable difference if the transmission coefficient $T$ can be found and physical quantities are compared at the same $T$.
We consider a uniform bias
$E_{L}=\mu_{B}/2$ on the left half and, similarly, $E_{R}=-\mu_{B}/2$ on the
right half. By making the two lattices on both sides semi-infinite,
they behave as the two reservoirs with different electrochemical potentials. The hopping coefficient is denoted by $\bar{t}$
and the unit of time is $t_{0}=\hbar/\bar{t}$. We set the electric
charge $e\equiv1$ and $\hbar\equiv1$. The length is measured in units of the lattice constant.

The Green's function of the left (right) semi-infinite chain can be derived using recursive relations, which lead to \cite{DNAspintronics}
$G_{L(R)}(E)=1/[E-E_{L(R)}-\Sigma_{L(R)}(E)]$,
where $\Sigma_{L(R)}=(1/2)\left[E-E_{L(R)}-i\sqrt{4\bar{t}^{2}-(E-E_{L(R)})^{2}}\right]$.
The retarded Green's function of the junction is $G(E)=1/[E-E_{C}-\Sigma_{CL}-\Sigma_{CR}]$,
where $\Sigma_{CL(CR)}=V_{CL(CR)}^{2}G_{L(R)}(E)$ and $V_{CL(CR)}$ is
the coupling to the left (right) chain \cite{DiVentrabook,GEnote}.
The current (including both spins) is \cite{DiVentrabook}
\begin{equation}
I=\frac{1}{\pi}\int_{-\infty}^{\infty}dE(f_{L}-f_{R})T(E)=\frac{1}{\pi}\int_{-\frac{\mu_{B}}{2}}^{\frac{\mu_{B}}{2}}dET(E),\label{eq:Ianalytic}
\end{equation}
where the reservoirs are taken to be at zero temperature, as we will throughout this work.
The transmission coefficient is
\begin{equation}
T(E)=\Gamma_{L}\Gamma_{R}|G(E)|^{2},\label{eq:TE}
\end{equation}
 where $f_{L(R)}$ denotes the density distribution of the left (right) chain, i.e.,  the Fermi-Dirac distribution function,
  and $\Gamma_{L (R)}=-2\mbox{Im}\Sigma_{CL (CR)}$.

For a uniform chain with $\bar{t}^{\prime}=\bar{t}$, $V_{CL(CR)}=\bar{t}$.
After some algebra, the current is given by
\begin{equation}\label{eq:LI_equal}
I=\frac{1}{\pi}\int_{-\frac{\mu_{B}}{2}}^{\frac{\mu_{B}}{2}}\frac{4g_{L}g_{R}dE}{\mu_{B}^{2}+(g_{L}+g_{R})^{2}},
\end{equation}
where $g_{L(R)}=\sqrt{4\bar{t}^{2}-(E-E_{L(R)})^{2}}$.
To the leading order of $\mu_{B}$,
Eq.~\eqref{eq:LI_equal} gives $I  \simeq \mu_B \bar{t}/\pi$.
Moreover, it can be shown that $T(E=0)\rightarrow1$ as $\mu_{B}\rightarrow0$.

For the weak-link case, if we take the last site of the left chain
as the central site, $V_{CL}=\bar{t}$, $V_{CR}=\bar{t}^{\prime}$,
and $E_{C}=E_{L}$. The current is
\begin{equation}\label{eq:LI_tp}
I=\frac{1}{\pi}\int_{-\frac{\mu_{B}}{2}}^{\frac{\mu_{B}}{2}}\frac{4g^{2}g_{L}g_{R}dE}{[(E-E_L)-g^{2} (E-E_R)]^2+(g_L+ g^{2}g_R)^2},
\end{equation}
where $g\equiv(\bar{t}^{\prime}/\bar{t})$.
When $g\ll 1$, to the leading order of $g$ and then to the leading order of $\mu_B$, one obtains $I  \simeq 4\mu_{B}g^2\bar{t}/\pi$.
For the central-site case, $V_{CL}=V_{CR}=\bar{t}$
and $E_{C}$ can be tuned.
The current is
\begin{equation}\label{eq:LI_EC}
I=\frac{1}{\pi}\int_{-\frac{\mu_{B}}{2}}^{\frac{\mu_{B}}{2}}\frac{4g_{L}g_{R}dE}{(E_L+E_R-2E_C)^2+(g_L+ g_R)^2}.
\end{equation}
Figure~\ref{fig:TE_2cases} shows $T(E)$, which is symmetric about $E=0$,
for both cases with selected parameters.

\section{Micro-canonical formalism}\label{sec:MCF}
In the micro-canonical approach to quantum transport~\cite{micro}, one considers a  finite system (say
two electrodes and a junction) and a finite number of particles with Hamiltonian $H$. The system is  prepared in an initial state $|\Psi_0\rangle$ which is an eigenstate of some Hamiltonian $H_{0} \neq H$. From a physical point of view this initial state may represent, e.g., a charge, particle, or energy imbalance between the two finite electrodes
that sandwich the junction.
The system is then left to evolve from this initial condition under the dynamics of $H$, and
the average current across some surface or any other observable is monitored in time. The dynamics considered here may be considered as quantum quenches \cite{Meisner09,Polkovnikov11}. Note that, even if we assume the
two electrodes biased as in the Landauer formalism, in this closed-system approach it is not at all obvious that
the average current establishes any (quasi-)steady state in the course of time~\cite{micro,DiVentrabook,Beria13}.

\subsection{Implementation of the MCF}
We adopt the implementation of the micro-canonical formalism as discussed in Refs.~\cite{MCFshort,MCF_TD},
which is an extension of the scheme proposed in Ref.~\cite{Bushong05}.
One advantage of this extended scheme is that the dynamics of particle
density fluctuations, entanglement entropy, and density distributions
can be easily monitored. We consider a one-dimensional Hamiltonian $H=H_{L}+H_{R}+H_{C}$,
where $H_{L/R}$ is a lattice of $N/2$ sites. The system is filled
with $N/2$ two-component fermions (with equal number in each species).
In the tight-binding approximation we choose
\begin{equation}
H_{L/R}=-\bar{t}\sum_{\langle ij\rangle,L/R}c_{i}^{\dagger}c_{j}+E_{L/R}\sum_{i\in L/R}c_{i}^{\dagger}c_{i}.
\end{equation}
 Here $\langle ij\rangle$ denotes nearest-neighbor pairs and we suppress
the spin index, and explicitly state where a summation over the spin
is performed in our results. Here we only consider quadratic Hamiltonians. In the presence of other interaction terms, one may need to consider approximate methods \cite{int_induced}.

We consider two possible ways to set the system out of equilibrium. In the first scenario the system is initially prepared in the ground
state of the unbiased Hamiltonian $H_{0}$ with $E_{L}=E_{R}=0$ and
then it evolves according to a biased Hamiltonian $H$. All conclusions in this work remain unchanged if we instead prepare the system
in the ground state of the biased $H$ and then let it evolve according to the unbiased $H_{0}$. In other words, there is a correspondence
between a particle imbalance and an energy imbalance for the systems we consider. In the second scenario two initially disconnected lattices are connected, where one can not swapped the roles of $H$ and $H_0$. We remark that the first scenario is closely related to the studies of Ref.~\cite{Atomtronics07} where photons are introduced to adjust the onsite energy of atoms in certain parts of the lattice. The second scenario is relevant to the case where an optical barrier separating the lattice into two parts is lifted \cite{MCF_TD}.

For the weak-link case $H_{C}=-\bar{t}^{\prime}(c_{N/2}^{\dagger}c_{N/2+1}+c_{N/2+1}^{\dagger}c_{N/2})$,
where $0\le\bar{t}^{\prime}\le\bar{t}$, while for the central-site
case $H_{C}=E_{C}c_{N/2+1}^{\dagger}c_{N/2+1}-\bar{t}(c_{N/2}^{\dagger}c_{N/2+1}+c_{N/2+1}^{\dagger}c_{N/2})-\bar{t}(c_{N/2+1}^{\dagger}c_{N/2+2}+c_{N/2+2}^{\dagger}c_{N/2+1})$.
For time $t<0$, $E_{L/R}=0$ and the system is in the ground state
of $H_{0}$. For $t>0$ we set $E_{L}=\mu_{B}/2$ and $E_{R}=-\mu_{B}/2$
and let the system evolve. Figure~\ref{fig:Schematic}
illustrates this process for the weak-link case. A uniform chain with
$\bar{t}^{\prime}=\bar{t}$ in the weak-link case has been shown to
have a quasi steady-state current (QSSC) at a small bias \cite{Bushong05}
for a system as small as $N=60$. The QSSC is defined as a plateau
in the current as a function of $t$ and it usually spans the range
$(N/4)t_{0}\le t\le(N/2)t_{0}$. In the thermodynamic limit with finite
filling, the QSSC becomes a steady current \cite{MCF_TD}. In contrast,
noninteracting bosons in their ground state do not support a QSSC
\cite{MCFshort,MCF_TD}. The dependence of the magnitude of the QSSC on the initial filling was discussed in Refs.~\cite{MCFshort,MCF_TD} and here we consider the case with $N_p/N=1/2$, where $N_p$ denotes the number of particles in the system, unless specified otherwise.

To gain more insight into the dynamics of the system, we write down
the correlation matrix $C(t)$ with elements $c_{ij}(t)=\langle GS_{0}|c_{i}^{\dagger}(t)c_{j}(t)|GS_{0}\rangle$,
where $|GS_{0}\rangle$ denotes the ground state of $H_{0}$, and
derive the current and entanglement entropy from it. One can use unitary
transformations $c_{j}=\sum_{k}(U_{0})_{jk}a_{k}$ and $c_{j}=\sum_{k}(U_{e})_{jk}d_{k}$
to rewrite $H_{0}$ and $H$ as
\begin{equation}
H_{0}=\sum_{k}\epsilon_{k}^{0}a_{k}^{\dagger}a_{k};\mbox{ }H=\sum_{p}\epsilon_{p}^{e}d_{p}^{\dagger}d_{p}.
\end{equation}
 Here $\epsilon_{k}^{0}$ and $\epsilon_{p}^{e}$ are the energy spectra
of $H_{0}$ and $H$, respectively. The initial state is then
$|GS_{0}\rangle=(\Pi_{k=1}^{N/2}a_{k}^{\dagger})|0\rangle$, where
$|0\rangle$ is the vacuum. From the equation of motion $i(dc_{j}(t)/dt)=[c_{j}(t),H]$
it follows $c_{j}(t)=\sum_{p}(U_{e})_{jp}d_{p}(0)\exp(-i\epsilon_{p}^{e}t)$.
The initial correlation functions are $\langle GS_{0}|a_{k}^{\dagger}(0)a_{k^{\prime}}(0)|GS_{0}\rangle=\theta(N/2-k)\delta_{k,k^{\prime}}$
since fermions occupy all states below the Fermi energy, where $\theta(N/2-k)$
is $1$ if $k\le N/2$, and $0$ otherwise. Then it follows
\begin{eqnarray}
c_{ij}(t) & = & \sum_{p,p^{\prime}=1}^{N}(U_{e}^{\dagger})_{pi}(U_{e})_{jp^{\prime}}D_{pp^{\prime}}(0)e^{i(\epsilon_{p}^{e}-\epsilon_{p^{\prime}}^{e})t};\label{eq:cij}\\
D_{pp^{\prime}}(0) & = & \sum_{m,m^{\prime}=1}^{N}\sum_{k=1}^{N/2}(U_{e}^{\dagger})_{p^{\prime}m^{\prime}}(U_{0})_{m^{\prime}k}(U_{0}^{\dagger})_{km}(U_{e})_{mp}.\nonumber
\end{eqnarray}
 Here $D_{pp^{\prime}}(0)\equiv\langle GS_{0}|d_{p}^{\dagger}(0)d_{p^{\prime}}(0)|GS_{0}\rangle$.

\subsection{Current, entanglement entropy, and particle fluctuations}
The current flowing from left to right for one species is
$I=-\langle d\hat{N}_{L}(t)/dt\rangle$, where $\hat{N}_{L}(t)=\sum_{i=1}^{N/2}c_{i}^{\dagger}(t)c_{i}(t)$.
It can be shown that for the Hamiltonian considered here, $I=4\bar{t}^{\prime}\mbox{Im}\{c_{(N/2),(N/2+1)}(t)\}$,
where a factor of $2$ for the two spin components is included. This is equivalent
to the expectation value of the current operator $\hat{I}=-i\bar{t}^{\prime}(c_{N/2}^{\dagger}c_{N/2+1}-c_{N/2+1}^{\dagger}c_{N/2})$.
The MCF can be generalized to include finite-temperature effects in the initial state \cite{MCFshort},
but here we focus on the ground state.

\begin{figure}
\includegraphics[clip,width=3.4in]{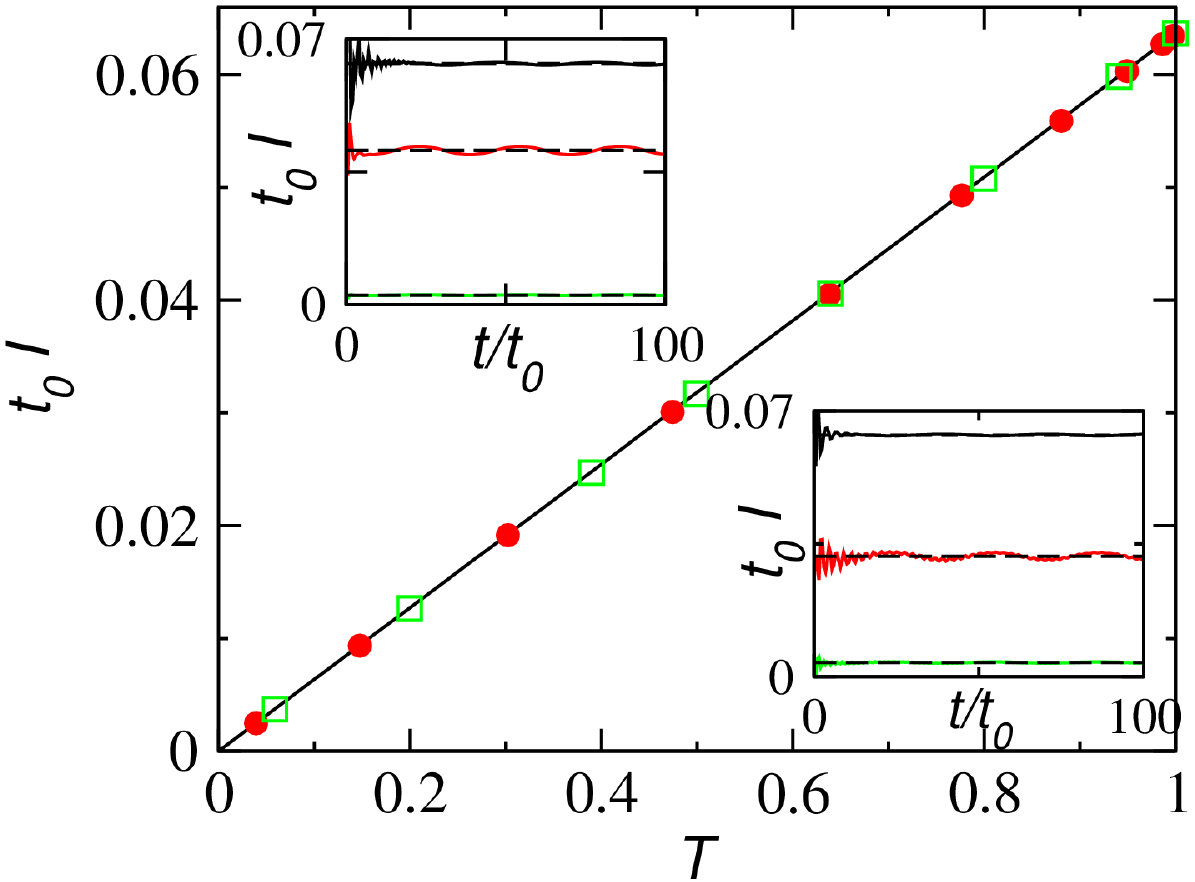} \caption{(Color online) The current from the Landauer formula for the weak-link case, Eq.~\eqref{eq:LI_tp}, (black
line) and the currents in the quasi-steady states of the micro-canonical simulations for the weak-link case (red circles)
and the central-site case (green squares) as a function of the transmission
coefficient $T=T(E=0)$. Insets: Currents as a function of time from
the micro-canonical simulations (solid lines; the dashed lines represent the Landauer value).
The upper (lower) one corresponds to the weak-link (central-site)
case. From top to bottom for the upper inset: $\bar{t}^{\prime}/\bar{t}=1.0,0.5,0.1$
and for the lower inset: $E_{C}/\bar{t}=0,2,8$. Here $\mu_{B}=0.2\bar{t}$
and $N=512$.}
\label{fig:Im_IL}
\end{figure}
Figure~\ref{fig:Im_IL} compares the current predicted by the Landauer formula for the weak-link case, Eq.~\eqref{eq:LI_tp} to the simulations using the MCF for the weak-link as well as the central-site cases with $\mu_{B}=0.2\bar{t}$. In the limit where $\mu_{B}\rightarrow 0$, the Landauer formulas for the central-site case, Eq.~\eqref{eq:LI_EC}, produces results that fully agree with the results from the weak-link case, Eq.~\eqref{eq:LI_tp}. When $\mu_{B}$ is finite, the two cases differ by a negligible amount due to the slightly different $T(E)$. One can see that the currents from the MCF agree well with that from the Landauer formula. 

The entanglement entropy between the left and right halves, $s$, for one species at time $t$ can be evaluated
as follows \cite{KlichLevitov}. We define a $(N/2)\times(N/2)$ matrix
$M=P_{L}C(t)P_{L}$ with elements $M_{ij}$, where the projection
operator $P_{L}=diag({\bf 1}_{N/2},{\bf 0}_{N/2})$. Then the entanglement
entropy can be obtained from the expression
\begin{equation}
s=-\mbox{Tr}[M\log M+(1-M)\log(1-M)].\label{eq:sdef}
\end{equation}
The Hermitian matrix $M_{ij}$ has eigenvalues $v_{i}$, $i=1\cdots N/2$.
Then
\begin{equation}
s(t)=\sum_{i=1}^{N/2}[-v_{i}\log(v_{i})-(1-v_{i})\log(1-v_{i})].
\end{equation}
We use $\log$ base $2$, as is convention.
This expression may be further simplified by using approximations
from the semi-classical FCS \cite{KlichLevitov} and will be discussed
later on. The noninteracting fermions studied here may be regarded as a limiting case of
a XXZ spin chain, whose entanglement entropy (due to the dynamics of magnetization) has been studied in
Ref.~\cite{Sabetta13}.

Now we derive the full quantum-mechanical expressions for the equal-time
number fluctuations of the left half lattice. Let $\hat{n}_{i}=c_{i}^{\dagger}c_{i}$
and $\hat{N}_{L}=\sum_{i=1}^{N/2}\hat{n}_{i}$. Then the number of particles
in the left part is $N_{L}=\langle\hat{N}_{L}\rangle=\sum_{i=1}^{N/2}c_{ii}$.
We define the equal-time number fluctuations of the left half as
\begin{eqnarray}
\Delta N_{L}^{2}=\langle(\hat{N}_{L}-N_{L})^{2}\rangle=\langle\hat{N}_{L}^{2}\rangle-N_{L}^{2}.\label{eq:dN2}
\end{eqnarray}
 The moments of $\hat{N}_{L}$ can be obtained from
\begin{eqnarray}
\langle\hat{N}_{L}^{2}\rangle & = & \sum_{i=1}^{N/2}\langle\hat{n}_{i}^{2}\rangle+2\sum_{i<j}^{N/2}\langle\hat{n}_{i}\hat{n}_{j}\rangle
\end{eqnarray}
 From Wick's theorem or exact calculations, $\langle\hat{n}_{i}^{\alpha}\rangle=\langle\hat{n}_{i}\rangle=n_{i}$
for all positive integer $\alpha$, where $n_{i}=c_{ii}$. The other
correlation functions can be obtained from Wick's theorem so that
\begin{eqnarray}
\langle\hat{n}_{i}\hat{n}_{j}\rangle & = & n_{i}n_{j}-|\langle c_{i}^{\dagger}c_{j}\rangle|^{2}.
\end{eqnarray}

\section{Spatial resolution of the current in MCF}\label{sec:MCFspatial}
We stress an important feature of the MCF formalism. One can see from
Eq.~\eqref{eq:cij} and its context that MCF monitors the dynamics
in both energy basis and real space. In contrast, the Landauer formalism
as shown in Eq.~\eqref{eq:Ianalytic} only reveals information in
the energy basis. The ability of the MCF to trace the dynamics in real
space allows us to address a crucial question: How do particles
from different sites contribute to the current?

To clearly demonstrate the importance of the information from the dynamics in real space, we consider a simplified
initial condition where $N$ lattice sites are divided into the left
$N/2$ sites and the right $N/2$ sites with each left site occupied
by one fermion and each right site empty. We consider a uniform lattice
here with a tunneling coefficient $\bar{t}$. The corresponding
correlation matrix is $c_{ij}(t=0)=\delta_{ij}$ if $1\le i,j\le(N/2)$
and zero otherwise. Eq.~\eqref{eq:cij} becomes
\begin{equation}
c_{ij}(t)=\sum_{m=1}^{N/2}\sum_{p,p^{\prime}=1}^{N}(U_{e}^{\dagger})_{pi}(U_{e})_{jp\prime}(U_{e}^{\dagger})_{mp}(U_{e})_{p^{\prime}m}e^{i(\epsilon_{p}^{e}-\epsilon_{p^{\prime}}^{e})t}
\end{equation}
 One important insight from this expression is that the index $m$
traces the contribution from the initially filled $m$-th site on
the left. Therefore in the current $I=-2\bar{t}\mbox{Im}(c_{N/2,N/2+1})$
it is meaningful to discuss where does the current come from as time
evolves.

This simplified case, despite its compactness and clarity, is relevant
to several situations realizable in experiments. Two potential
examples are: (1) initially a large step-function bias is applied
to a nanowire with a small energy bandwidth so that all mobile particles
are driven to the left half and then the bias is removed to allow
a current to flow, and (2) ultra-cold atoms are loaded in an optical
lattice so that there is one atom per lattice site. Then a focus laser
beam excites the atoms on the right half lattice so that they leave
the lattice and create a vacuum region. The atoms on the filled left
part will then flow to the right and build a current. Thus the physics
of this simplified case is relevant to both our deeper understanding
of transport phenomena and advances in experiments.

\begin{figure}
\includegraphics[clip,width=3.4in]{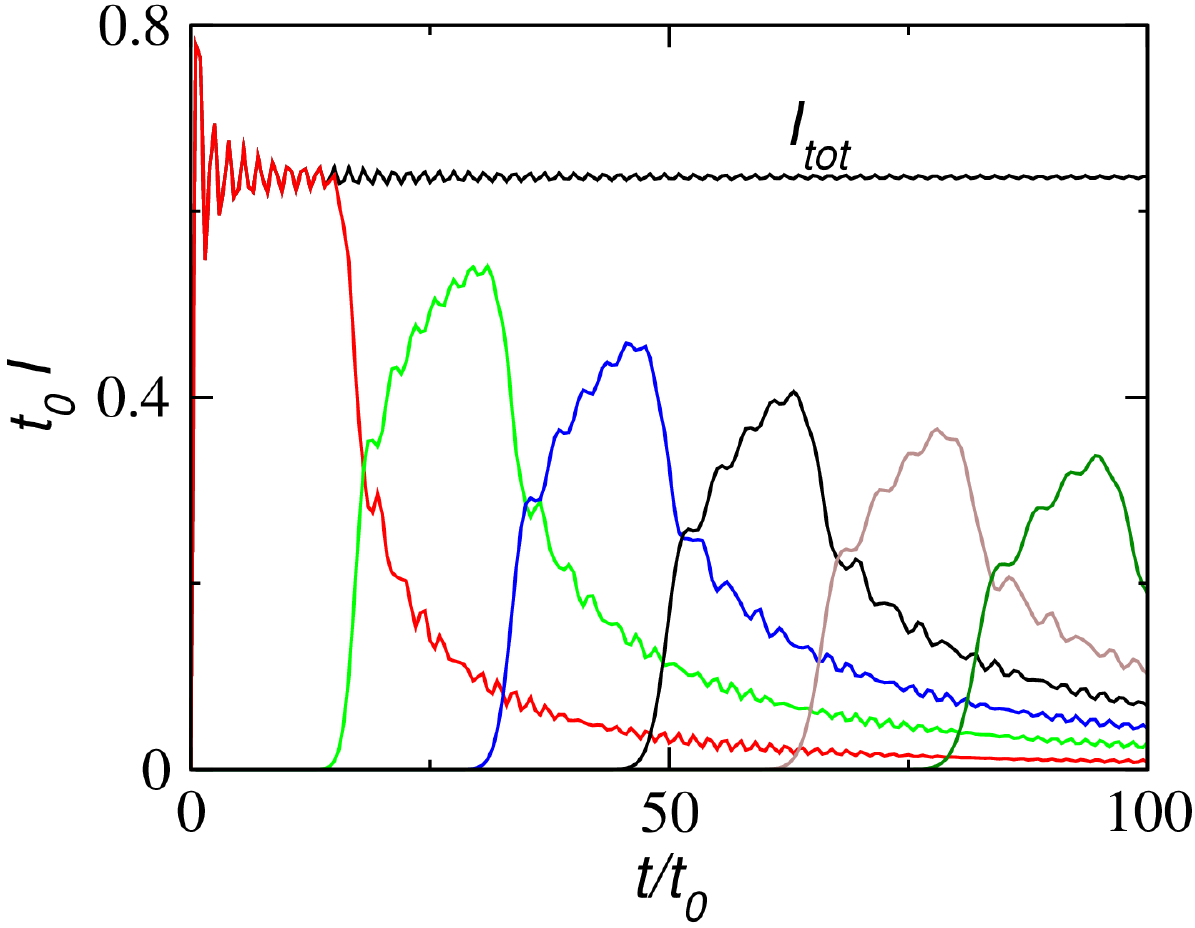}
\caption{(Color online) Spatial decomposition of the contribution to the current.
The black line labeled $I_{tot}$ shows
the current from an $N=512$ lattice with the left half initially
filled with one fermion per site. We plot the contributions from sections
of $32$ sites each to the left of the middle of the whole lattice
and the corresponding currents show up in bursts. The bursts, from
left to right on the plot, correspond to the current from the first,
second, ..., sixth sections of $32$ sites to the left away from the
middle (the $256$-th site). \label{fig:I_decomp}}
\end{figure}

Figure~\ref{fig:I_decomp} shows the total current of this case with
$N=512$ and clearly there is a quasi steady-state current. When we
determine the contributions from each section of $32$ lattices sites
to the left of the middle ($256$-th site), each contribution comes
in a burst following the previous burst from the section to its right.
Thus the burst from the section of the $225$-th site to the $256$-th
site crosses the middle first, followed by the burst from the section
of the $193$-th site to the $224$-th sites, and so on, with each burst
having a decaying tail. This succession of bursts gives a physical
justification of the reason the semi-classical distribution assumed in FCS is binomial, and why other distributions
can be excluded (see below). Each burst peak plus all the tails
from previous bursts add up to maintain the observed quasi steady-state
current. The MCF formalism thus provides more insights into how a
quasi steady-state current forms and this is certainly beyond the
scope of the Landauer's formalism.
Since some spin chain problems can be mapped to fermions in 1D,
our study is relevant to the dynamics of magnetization in these cases as well \cite{Antal99}.

\section{Semi-classical FCS formalism}\label{sec:FCS}
For two 1D non-interacting fermionic systems connected by a central
barrier, it has been proposed \cite{KlichLevitov} that an expression
for the entanglement entropy can be derived from FCS assuming a binomial
distribution of the transmitted particle number. In linear response, it has the form
\begin{equation}
\frac{\Delta s}{\Delta t}=-2\frac{\mu_{B}}{h}[T\log T+(1-T)\log(1-T)].\label{eq:Klich_s}
\end{equation}
Here, $T$ is the transmission coefficient at the Fermi
energy. The second moment of transmitted particle numbers, $\mathcal{C}_{2}$,
is important because it may be inferred from shot-noise measurements.
Moreover, the spectrum of current fluctuations through the barrier,
$P_{sn}$, is related to $\mathcal{C}_{2}$ by $P_{sn}=\mathcal{C}_{2}/t$.
Refs.~\cite{KlichLevitov} gives the prediction for $P_{sn}$:
\begin{equation}
P_{sn}=\frac{\mathcal{C}_{2}}{t}=\frac{2\mu_{B}}{h}T(1-T).\label{eq:shotnoise}
\end{equation}
 We will briefly review the derivations for these expressions.

In a semi-classical description, the second moment of transmitted
particle numbers, $\mathcal{C}_{2}$, is equivalent to the number
fluctuations of the left half of the system if the number of particles
are conserved. This can be understood as follows. Let us assume that at time
$t$ there are $N_{L0}$ particles on the left. At time $t+\Delta t$,
if there are $N_{T}$ particles passing through the barrier, the total
number of particles on the left becomes $N_{L}=N_{L0}-N_{T}$. When
$N_{L0}$ is treated as a number, one has $\mathcal{C}_{2}=\langle N_{T}^{2}\rangle-\langle N_{T}\rangle^{2}=\langle N_{L}^{2}\rangle-\langle N_{L}\rangle^{2}=\Delta N_{L}^{2}$.

In a fully quantum-mechanical description, however, $N_{L0}$ is
an operator and the cross-correlation $\langle\hat{N}_{L0}\hat{N}_{T}\rangle\neq\langle\hat{N}_{L0}\rangle\langle\hat{N}_{T}\rangle$
may introduce corrections to the expression. In the micro-canonical
formalism, the fully quantum-mechanical equal-time number fluctuations,
$\Delta N_{L}^{2}$, can be monitored. We will compare this with the
prediction of $\mathcal{C}_{2}$ from the semi-classical formula Eq.~(\ref{eq:shotnoise})
and see how important the quantum corrections are.

We summarize how the moments and entanglement entropy can be evaluated
from semi-classical FSC \cite{KlichLevitov}. The characteristic function
(CF) of transmission of fermions of one species is $\chi(\lambda)=\sum_{n=-\infty}^{\infty}P_{n}e^{i\lambda n}$,
where $P_{n}$ is the probability of $n$ fermions being transmitted. In terms
of cumulants of FCS,
\begin{equation}
\log\chi(\lambda)=\sum_{m=1}^{\infty}\frac{(i\lambda)^{m}}{m!}\mathcal{C}_{m}.
\end{equation}
 Importantly, the generating function is shown to be \cite{KlichLevitov}
\begin{equation}
\chi(\lambda)=\det\left((1-M+Me^{i\lambda})e^{-i\lambda X}\right),
\end{equation}
 where $X=\exp(iHt)C(0)P_{L}\exp(-iHt)$ and $P_{L}$ is the
projection operator into the left-half lattice. Using $\det(AB)=\det(A)\det(B)$
and $\log\det(A)=Tr\log(A)$ one obtains
\begin{equation}
\log\chi(\lambda)=-i\lambda x+\log[\det(1-M+Me^{i\lambda})],\label{eq:logchi}
\end{equation}
 where $x=Tr(X)$ and $Tr$ denotes the trace. The matrix $M$ can
be diagonalized as $M=SD_{M}S^{\dagger}$, where $D_{M}=diag(v_{1},\cdots,v_{N/2})$
and $S$ is a unitary matrix. Then we get the final expression
\begin{equation}
\log\chi(\lambda)=-i\lambda x+\log\prod_{j=1}^{N/2}(1-v_{j}+v_{j}e^{i\lambda}).
\end{equation}
 The second cumulant can be obtained from
\begin{eqnarray}
\mathcal{C}_{2} & = & \frac{\partial^{2}\log\chi(\lambda)}{\partial(i\lambda)^{2}}\Big|_{\lambda\rightarrow0}=\sum_{j=1}^{N/2}(v_{j}-v_{j}^{2}).\label{eq:C2}
\end{eqnarray}

The entanglement entropy defined in Eq.~\eqref{eq:sdef} can be calculated
as
\begin{equation}
s=-\int_{0}^{1}dz\mu(z)[z\log z+(1-z)\log(1-z)].\label{eq:s_integral}
\end{equation}
 Here $z=1/(1-e^{i\lambda})$ and the spectral weight $\mu(z)$ is
given by
\begin{equation}
\mu(z)=\frac{1}{\pi}\mbox{Im}\partial_{z}\log\chi(z-i0^{+}).
\end{equation}

The CF of a binomial distribution with a transmitted probability $T$
is $\chi(\lambda)=(1-T+Te^{i\lambda})^{\mathcal{N}}=(1-T/z)^{\mathcal{N}}$,
where $\mathcal{N}=2\mu_{B}\Delta t/h$ is the flux of incoming particles.
The spectral weight is then
\begin{eqnarray}
\mu(z) & = & \frac{1}{\pi}\mbox{Im}\partial_{z}\log(1-\frac{T}{z-i0^{+}})\nonumber \\
 & = & \mathcal{N}\frac{T}{z}\delta(z-T)\nonumber \\
 & = & \mathcal{N}\delta(z-T).
\end{eqnarray}
 In this derivation we have used $1/(x-i0^{+})=P(1/x)+i\pi\delta(x)$,
$\delta(z(z-T))=(1/z)\delta(z-T)$, and $(T/z)\delta(z-T)=\delta(z-T)$,
where $P$ denotes the Cauchy principal value. Then the entanglement
entropy of Eq.~\eqref{eq:s_integral} leads to the expression of
Eq.~\eqref{eq:Klich_s}. A similar calculation using Eq.~\eqref{eq:C2}
gives the expression of Eq.~\eqref{eq:shotnoise}.

\section{Absence of memory effects for non-interacting systems}\label{sec:memory}
Before presenting a comparison of the MCF results with those of the Landauer formalism, we first investigate how sensitive the MCF results
are to the time-dependence of the switch-on of the bias. This is important because in the
Landauer formalism a steady state is assumed from the outset, while in the MCF a quasi-steady state develops in time
and therefore its magnitude can be dependent on initial conditions and transient behavior of the bias.

So far we
only considered a sudden quench so that $\mu_{B}$ is abruptly switched
to its full value. The MCF can be applied to other scenarios beyond a sudden quench. Here
we consider situations where $\mu_{B}$
is switched on at a finite rate and reaches its full value at time
$t_{m}$. Here, we focus on the weak-link case and one has to monitor
the dynamics of the correlation matrix by solving the equations of
motion
\begin{eqnarray}
i\frac{\partial\langle c_{i}^{\dagger}c_{j}\rangle}{\partial t} & = & X-\frac{\mu_{B}}{2}\langle c_{i}^{\dagger}c_{j}\rangle_{i\in L}+\frac{\mu_{B}}{2}\langle c_{i}^{\dagger}c_{j}\rangle_{j\in L}+\nonumber \\
 &  & \frac{\mu_{B}}{2}\langle c_{i}^{\dagger}c_{j}\rangle_{i\in R}-\frac{\mu_{B}}{2}\langle c_{i}^{\dagger}c_{j}\rangle_{j\in R}.\label{eq:fEOM}
\end{eqnarray}
 Here, $X\equiv[\bar{t}^{\prime}\delta_{i,N/2}+\bar{t}(1-\delta_{i,N/2})]\langle c_{i+1}^{\dagger}c_{j}\rangle+[\bar{t}^{\prime}\delta_{i,N/2+1}+\bar{t}(1-\delta_{i,N/2+1})]\langle c_{i-1}^{\dagger}c_{j}\rangle-[\bar{t}^{\prime}\delta_{j,N/2}+\bar{t}(1-\delta_{j,N/2})]\langle c_{i}^{\dagger}c_{j+1}\rangle-[\bar{t}^{\prime}\delta_{j,N/2+1}+\bar{t}(1-\delta_{j,N/2+1})]\langle c_{i}^{\dagger}c_{j-1}\rangle$.
The equations of motion are derived from $i(\partial\langle c_{i}^{\dagger}c_{j}\rangle/\partial t)=\langle[c_{i}^{\dagger},H]c_{j}\rangle+\langle c_{i}^{\dagger}[c_{j},H]\rangle$,
where $[\cdot,\cdot]$ denotes the commutator of the corresponding
operators. We assume that the dynamics of the two spins are identical
and the initial condition is the same as that in the sudden-quench
case.

\begin{figure}
\includegraphics[clip,width=3.4in]{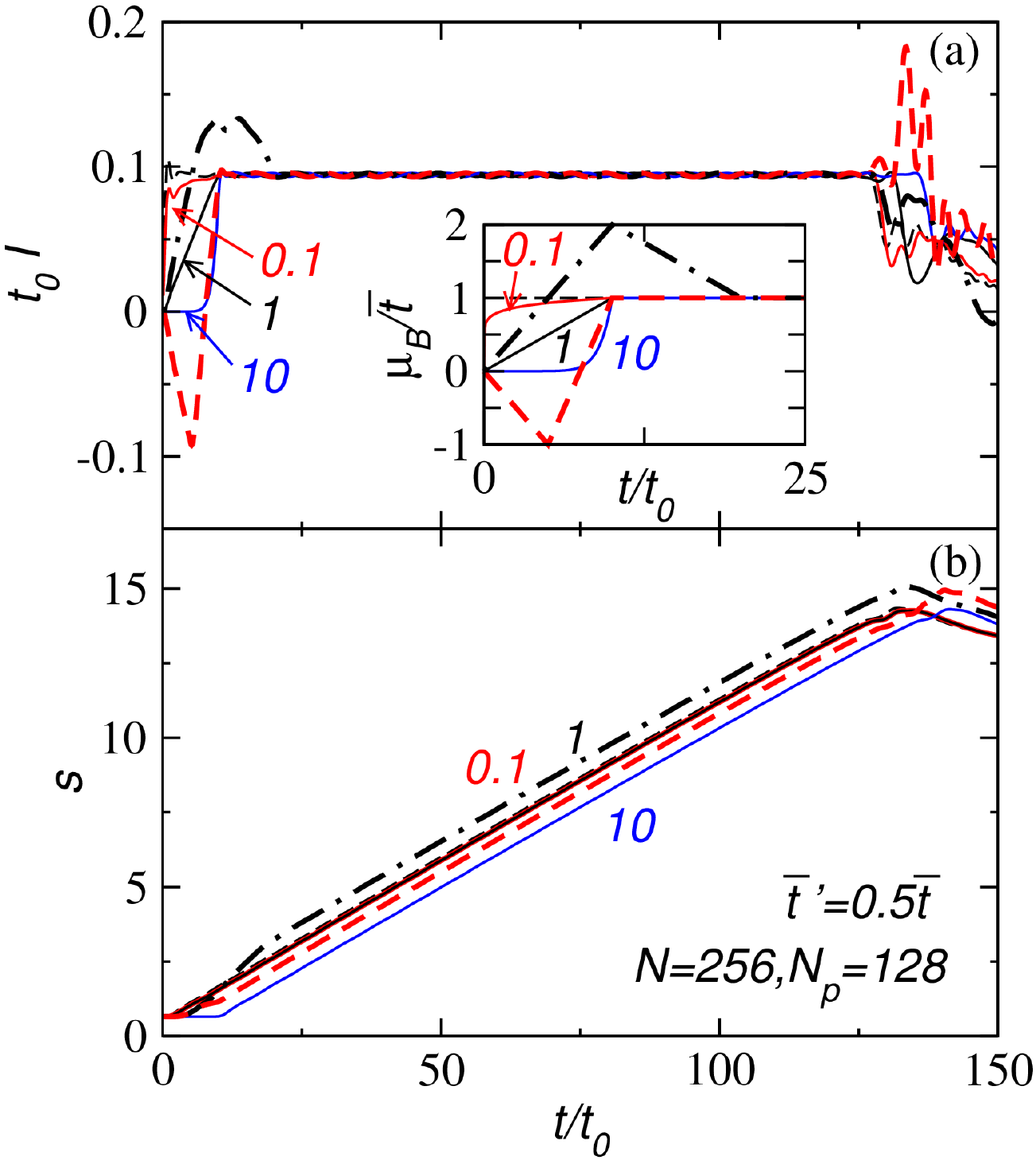}
\caption{(Color online) (a) Current and (b) entanglement entropy for different
time dependence of the bias $\mu_{B}=(t/t_{m})^{\alpha}\bar{t}$ for
$t<t_{m}$. Here $t_{m}=10t_{0}$, $\bar{t}^{\prime}/\bar{t}=0.5$,
$N=256$ and $N_{p}=128$. We show the results for $\alpha=0.1,1,10$
labeled next to each curve along with the results from a sudden quench
(dashed lines) and from a multi-step switching-on (dot-dash line).}
\label{fig:mem_vst}
\end{figure}

Figure~\ref{fig:mem_vst} shows the current and entanglement entropy
from different cases with $\mu_{B}(t)=(t/t_{m})^{\alpha}\bar{t}$
for $t<t_{m}$ and $\mu_{B}=\bar{t}$ for $t\ge t_{m}$. One can see
that despite different transient behaviors, the currents reach the
same magnitude when QSSCs emerge. Moreover, the slopes of the entanglement
entropy are also the same in the regime where QSSCs emerge. We find
the same conclusion when $t_{m}$ is varied. Importantly, one may
over-excite the system by tuning the bias {\it above} its final constant
value, and yet this spike does not affect the height of the QSSC or the
slope of the entanglement entropy as shown by the dot-dash lines in
Fig.~\ref{fig:mem_vst}.

Our observations then suggest that there is no observable memory effect
in the QSSC and entanglement entropy of non-interacting fermions driven
by a step-function bias because those observables are not sensitive
to the details of how the bias is turned on. However, the robustness of the
QSSC against different time dependencies of the switch-on of the bias
may not hold in the presence of interactions, and we leave this study for future work.

\begin{figure}
\includegraphics[clip,width=3.4in]{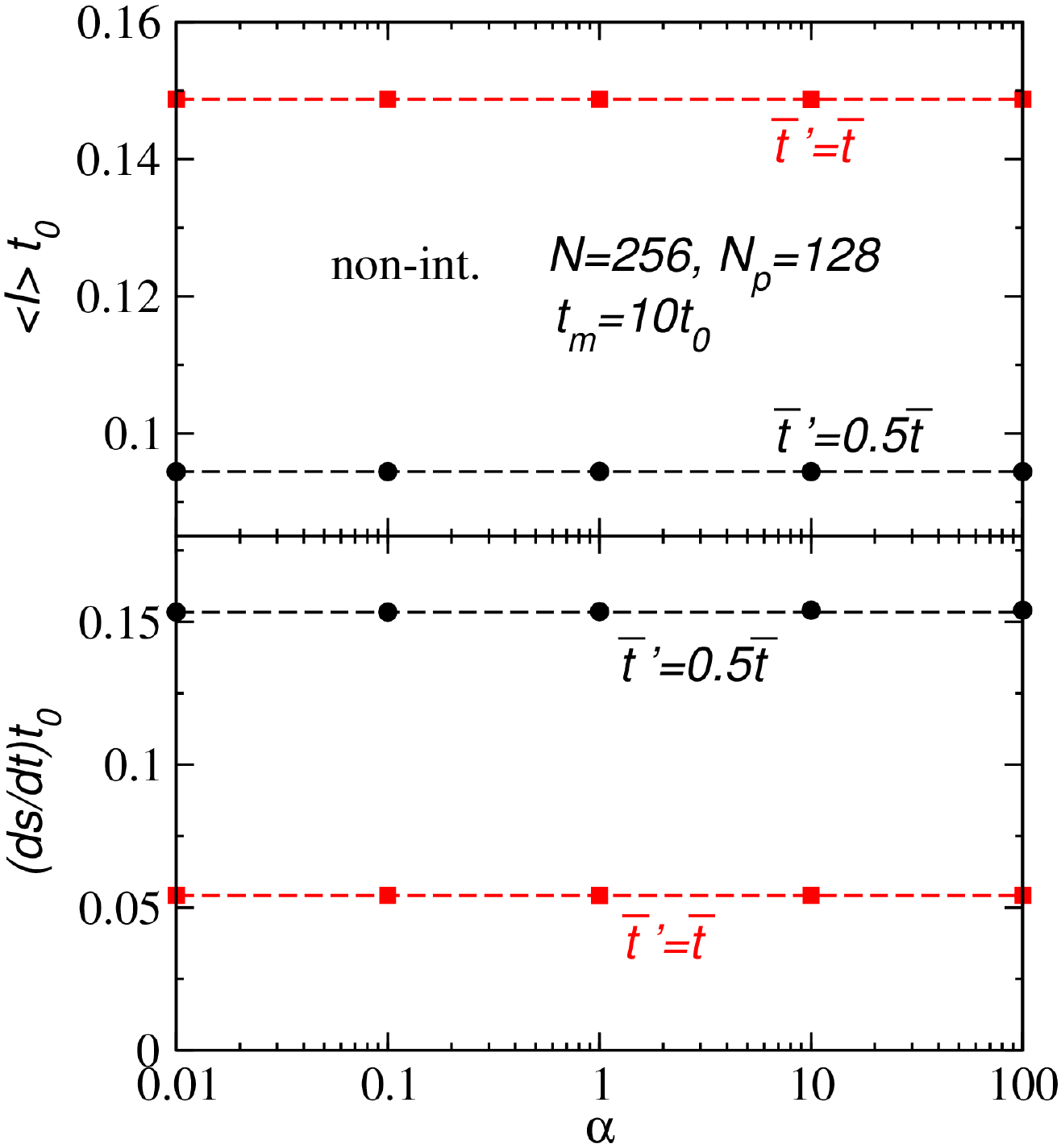} \caption{(Color online) (a) Averaged current (Eq.~\eqref{eq:Iavg}) and (b)
slope of the entanglement entropy for different time dependence of
the bias $\mu_{B}(t/t_{m})^{\alpha}\bar{t}$ for $t<t_{m}$. Here
$t_{m}=10t_{0}$, $\bar{t}^{\prime}/\bar{t}=0.5$ (black) and $\bar{t}^{\prime}/\bar{t}=1$
(red), $N=256$ and $N_{p}=128$. We show the results for $\alpha=0.01,0.1,1,10,100$
along with the results from a sudden quench (dashed lines).}

\label{fig:mem_IandS}
\end{figure}

Figure~\ref{fig:mem_IandS} shows the averaged current
\begin{equation}
\langle I\rangle=\frac{1}{50t_{0}}\int_{50t_{0}}^{100t_{0}}dtI(t)\label{eq:Iavg}
\end{equation}
 and the slope of $s$ in the region $50t_{0}\le t\le100t_{0}$ for
$\alpha=0.01,0.1,1,10,100$ along with the results from a sudden quench.
The results from those cases where the bias is turned on in a finite
time $t_{m}$ exhibit no observable deviation from the results from
the case of a sudden quench. We choose $t_{m}=10t_{0}$ and $N=256$
with $N_{p}=128$, but the conclusion holds for other parameters.
Thus in the following we focus on the sudden-quench case when we compare
the MCF and analytical formulas.

\section{Comparisons}\label{sec:comparison}
\begin{figure}
\includegraphics[clip,width=3.4in]{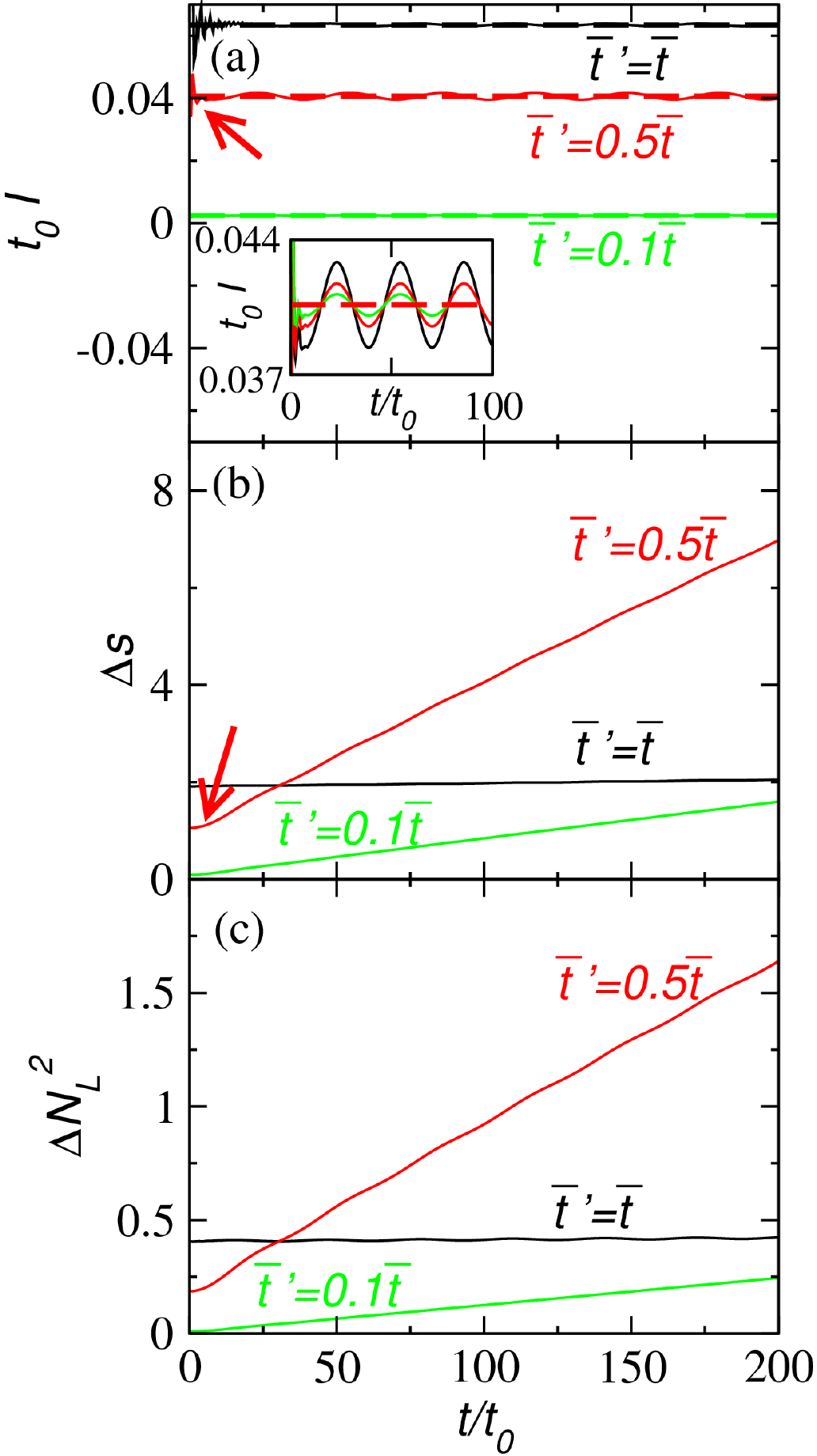} \caption{(Color online) (a) Current, (b) entanglement entropy, and (c) particle
number fluctuations of the weak-link case with $\bar{t}^{\prime}/\bar{t}=1,0.5,0.1$
(labeled next to the corresponding curve). The dashed lines in (a)
show the current predicted by Landauer formula, Eq.~(\ref{eq:Ianalytic}),
with the corresponding parameters. The arrows point to the transient regime where the current fluctuates in (a) and the entropy deviates from the linear dependence in (b) for the case with $\bar{t}^{\prime}/\bar{t}=0.5$. The inset shows $I$ for $N=256,512,1024$
(from the largest oscillation amplitude to the smallest one) and the
current predicted by Eq.~(\ref{eq:Ianalytic}) (dashed line) for
$\bar{t}^{\prime}/\bar{t}=0.5$. Here $\mu_{B}=0.2\bar{t}$
and $N=512$.}
\label{fig:sI}
\end{figure}

Figure~\ref{fig:sI} shows the current, entanglement entropy, and
number fluctuations of the weak-link case for selected values of $\bar{t}^{\prime}/\bar{t}$.
The currents clearly exhibit a quasi-steady state after a short transient
time. We emphasize again that the steady-state current results from the
quantum dynamics of the system and is not assumed {\it a priori}. The corresponding
steady-state currents calculated from the Landauer formula, Eq.~(\ref{eq:Ianalytic}),
are plotted on the same figure. The results from our simulations agree
well with the predictions from Landauer formula. This agreement is in line with
the observation that as the system approaches thermodynamic limit ($N\rightarrow\infty$ with finite filling), the microcanonical setup becomes indistinguishable from that in the Landauer formalism. For the case $\bar{t}^{\prime}=\bar{t}$,
we recover the quantized conductance, $G_{0}=I/\mu_{B}=2e^{2}/h$ (spin
included). As expected, in the presence of a barrier ($\bar{t}^{\prime}<\bar{t}$),
the conductance is smaller than the quantized conductance. The suppression of the current by a weak central link was also shown in Ref.~\cite{Waitingtime14}.
We also
test finite-size effects by comparing the currents from $N=256,512,1024$
with the prediction from Landauer formula in the inset of Fig.~\ref{fig:sI}.
One can see that while the oscillation amplitude decreases with increasing
system size, the average currents of the three different sizes all
agree well with the analytical result. For the central-site case we
found similar results.

When the link strength $\bar{t}^{\prime}$ or the central-site energy
$E_{C}$ is tuned, the transmission coefficient $T(E)$ changes accordingly.
Figure~\ref{fig:Im_IL} compares the quasi steady-state currents
from Landauer formula (black line) and from micro-canonical simulations
of the weak-link case (red circles) and the central-site case (green
squares) as a function of the transmission coefficient $T=T(E=0)$.
The three results agree well and this supports the expectation that the Landauer formalism provides reasonable predictions.
However, we will see that the agreement
does not hold when we study the distributions on the two sides of
the junction.

The entanglement entropy is expected to be linear in time and our results
support this claim. We found that the slope of $\Delta s=s-s(\mu_{B}=0)$
is proportional to $\mu_{B}$ as predicted in Eq.~\eqref{eq:Klich_s}. For different values of $\bar{t}^{\prime}/\bar{t}$,
we test the predictions from the two formulas. From
Fig.~\ref{fig:TE_2cases} we find that in the range $-\mu_{B}/2\le E\le\mu_{B}/2$
the variation of $T(E)$ is within $3\%$ for all cases we studied
so we take $T(E=0)$ as the transmission coefficient in our evaluation
of Eq.~\eqref{eq:Klich_s}.

\begin{figure}
  \includegraphics[width=3.4in,clip]{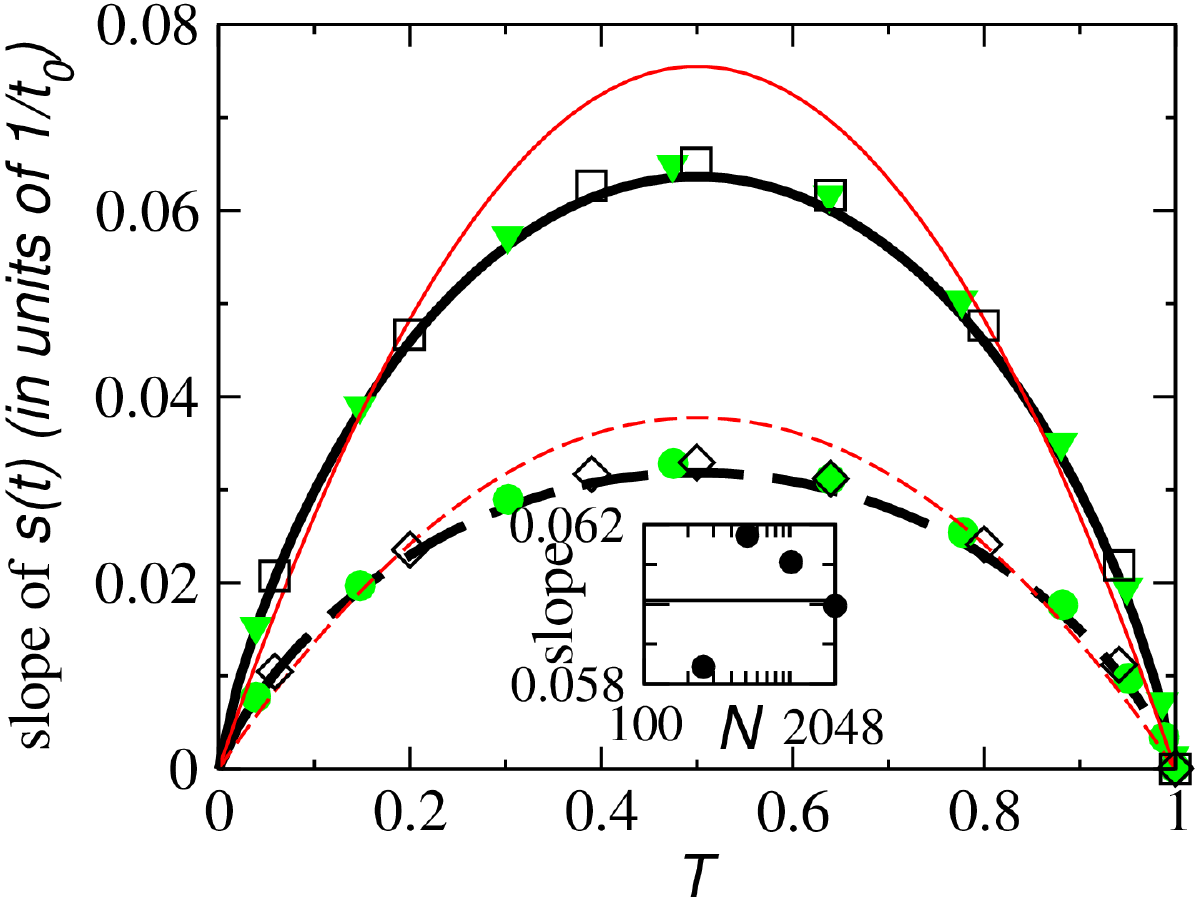}
  \caption{(Color online) Comparison of the slope of $s(t)$ from Eq.~\eqref{eq:Klich_s} (red) and simulations (symbols). Here the results for $\mu_B/\bar{t}=0.1$ are represented by the dashed line, circles (weak-link), and diamonds (central-site) while those for $\mu_B/\bar{t}=0.2$ are represented by the solid line, triangles (weak-link), and squares (central-site). We choose  $N=512$ and $T=T(E=0)$. The thin red solid and dashed lines shows the results for a Gaussian distribution, Eq.~\eqref{eq:Klich_Gaussian}, for $\mu_B/\bar{t}=0.2$ and $0.1$. Inset: The slopes (in units of $t_0^{-1}$) from different system sizes $N=256,512,1024,2048$ showing the convergence to the semi-classical value for $\bar{t}^{\prime}/\bar{t}=0.5$ (solid line).}
\label{fig:slope_comparison}
\end{figure}

The slope of the entanglement entropy from micro-canonical formalism
and the predictions from Eq.~\eqref{eq:Klich_s} are shown in
Figure~\ref{fig:slope_comparison} for $\mu_{B}/\bar{t}=0.1$ and
$0.2$. One can see that our results agree well with Eq.~\eqref{eq:Klich_s}
for all values of $T$ and this implies that the distribution of tunneling
particles may be approximated by a binomial form as assumed in Ref.~\cite{KlichLevitov}.
In the derivation of Eq.~\eqref{eq:Klich_s} and in our simulation, fermions
of different spins tunnel independently and do not generate spin-entangled
states. The entanglement entropy comes from the correlation of partially
tunneled and partially reflected wavefunctions of particles.

We notice that the transient time, which is defined as the initial
time interval during which the system has not reached a quasi-steady
state, seems to differ in $I(t)$ and $s(t)$ (as illustrated in Fig.~\ref{fig:sI}). During the transient
time, the currents fluctuate violently while the entropy exhibit a
downward bending. From our simulations we found that the transient
time for $s(t)$ is three times larger than that for $I(t)$ and this
relation seems to be insensitive to the system size.

To investigate how FCS depends on the underlying probability distribution, we study the behavior of Eq.~\eqref{eq:Klich_s} when a Gaussian (continuous) distribution is implemented.
To make connections with the original binomial distribution, we choose
the mean $m=\mathcal{N}T$ and the variance $\sigma^{2}=\mathcal{N}T(1-T)$
to match those of the binomial distribution. The CF is $\chi(\lambda)=\exp(im\lambda-\frac{1}{2}\sigma^{2}\lambda^{2})$.
A change of variable $z=1/(e^{i\lambda}-1)$ gives $\chi(z)=\exp(\mu\log(1-\frac{1}{z})+\frac{1}{2}\sigma^{2}[\log(1-\frac{1}{z})]^{2})$.
One then finds $\mbox{Im}\partial_{z}\log\chi(z)=\mbox{Im}[\frac{m}{z^{2}-z}+\frac{\sigma^{2}}{z^{2}-z}\log(1-\frac{1}{z})]$.
When one changes $z$ to $z-i0^{+}$ and uses the formula $1/(x-i0^{+})=P(1/x)+i\pi\delta(x)$,
the imaginary part of $1/(z^{2}-z)$ does not contribute to the integral
of $s$ because the delta function $\delta(z^{2}-z)$ can only be satisfied at $z=0,1$ but those points do not have finite $s$.
The only contribution in the spectral weight is thus $\mu(z)=\frac{1}{\pi}\frac{\sigma^{2}}{z^{2}-z}\mbox{Im}[\log(1-\frac{1}{z-i0^{+}})]$.
One can show that $\mbox{Im}[\log(1-\frac{1}{z-i0^{+}})]=\arg((z-1)/z)=-\pi$
(the choice of the sign will be clear in a moment) for $0<z<1$. Thus
the weight is $\mu(z)=-\frac{\sigma^{2}}{z^{2}-z}$ which is positive
for $0<z<1$. From $s=-\int_{0}^{1}dz\mu(z)[z\log z+(1-z)\log(1-z)]$
\cite{KlichLevitov} one gets $s=\alpha \sigma^{2}=\alpha(2\mu_{B}/h)T(1-T)\Delta t$.
Thus
\begin{equation} \label{eq:Klich_Gaussian}
\frac{\Delta s}{\Delta t}= \alpha \left(\frac{2\mu_B}{h}\right)T(1-T),
\end{equation}
where $\alpha \approx 3.3$ is a numerical factor.
In Figure~\ref{fig:slope_comparison} we show (in thin red lines) its values. It is clear that the data from the micro-canonical simulations can distinguish these distributions.

As the system size increases, the small oscillation on top of the linear increase of the entanglement entropy decreases. We found that this reduces the difference between the slope from fitting the results from the MCF and the slope predicted by the semi-classical FCS formalism. In the inset of Fig.~\ref{fig:slope_comparison} we show the slope from the MCF for $N=256,512,1024,2048$ with half-filling. One can see that as $N$ increases the agreement improves. However, optical lattices in real experiments are of limited sizes so one may expect observable finite-size effects in experimental results.

\begin{figure}
  \includegraphics[width=3.4in,clip]{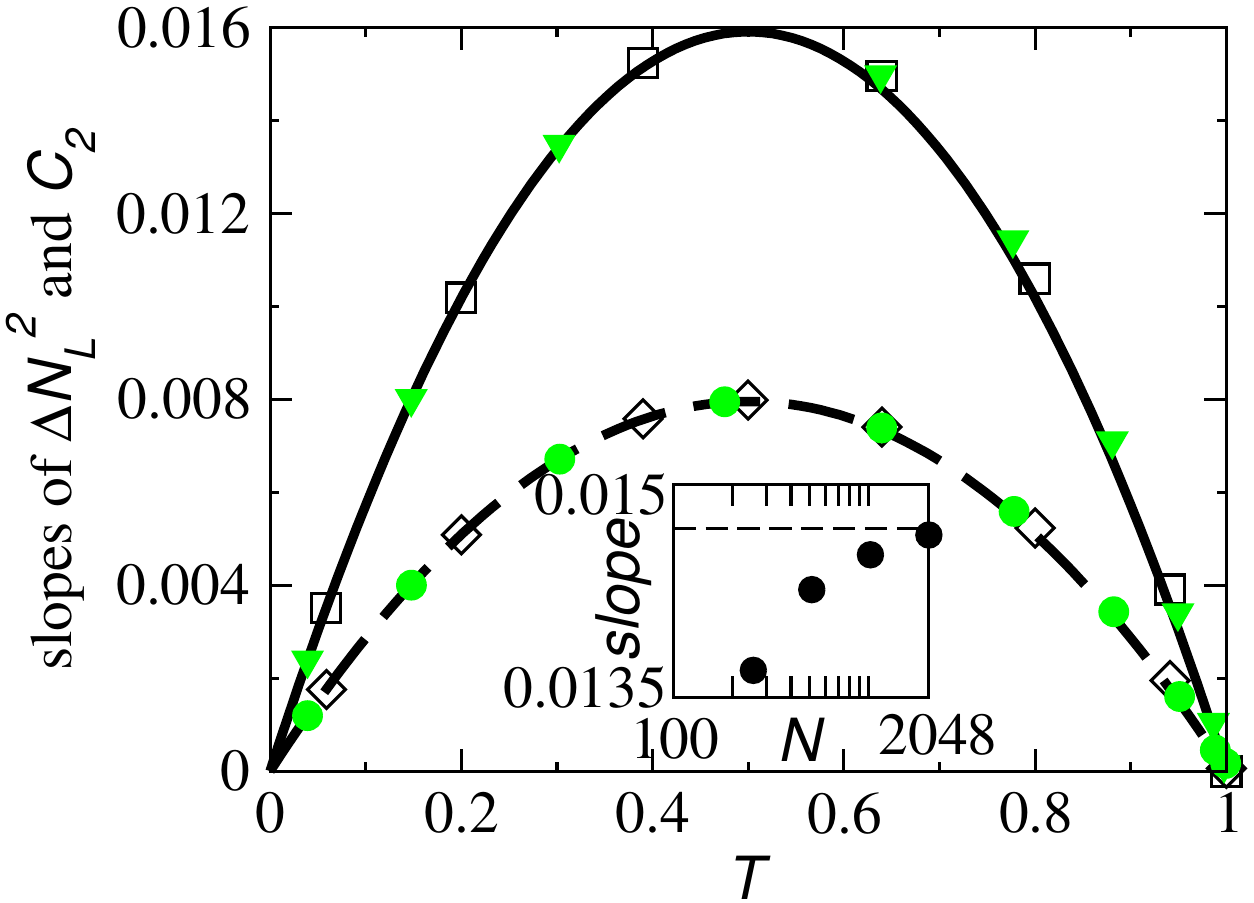}
  \caption{Comparison of the slopes (in units of $t_0^{-1}$) of $\Delta N_{L}^{2}$ (symbols) and $\mathcal{C}_{2}$ from Eq.~(\ref{eq:shotnoise}) (curves). The circles (weak-link), diamonds (central-site), and dashed line correspond to $\mu_B/\bar{t}=0.1$ while the triangles (weak-link), squares (central-site), and solid line correspond to $\mu_B/\bar{t}=0.2$. Here $N=512$ with half filling. Inset: The slope (in units of $t_0^{-1}$) of $\Delta N_{L}^{2}$ for $N=256,512,1024,2048$ at half filling with $\bar{t}^{\prime}/\bar{t}=0.5$. The dashed line shows the result from Eq.~\eqref{eq:shotnoise}.}
\label{fig:dNL2_cmp}
\end{figure}
Next we extract the slopes of $\Delta N_{L}^{2}$ and compare the
results with the slopes predicted from the semi-classical formula
of the second cumulant, Eq.~(\ref{eq:shotnoise}), in Figure~\ref{fig:dNL2_cmp}.
The slopes agree reasonably well, which implies that quantum
corrections to the semi-classical formula are insignificant.
Moreover,
we have compared the third and fourth moments with the quantum-mechanical
fluctuations of the corresponding order. The results from micro-canonical
simulations show observable deviations from those from semi-classical
FCS in the fourth order but not in the third order.
The slight difference between our results and the results
from semiclassical FCS in Fig.~\ref{fig:dNL2_cmp} is due to finite-size effects. We
have checked our results for larger system sizes and the result converges to the FCS prediction,
as shown in the inset of Fig.~\ref{fig:dNL2_cmp}.

\begin{figure}
\includegraphics[clip,width=3.4in]{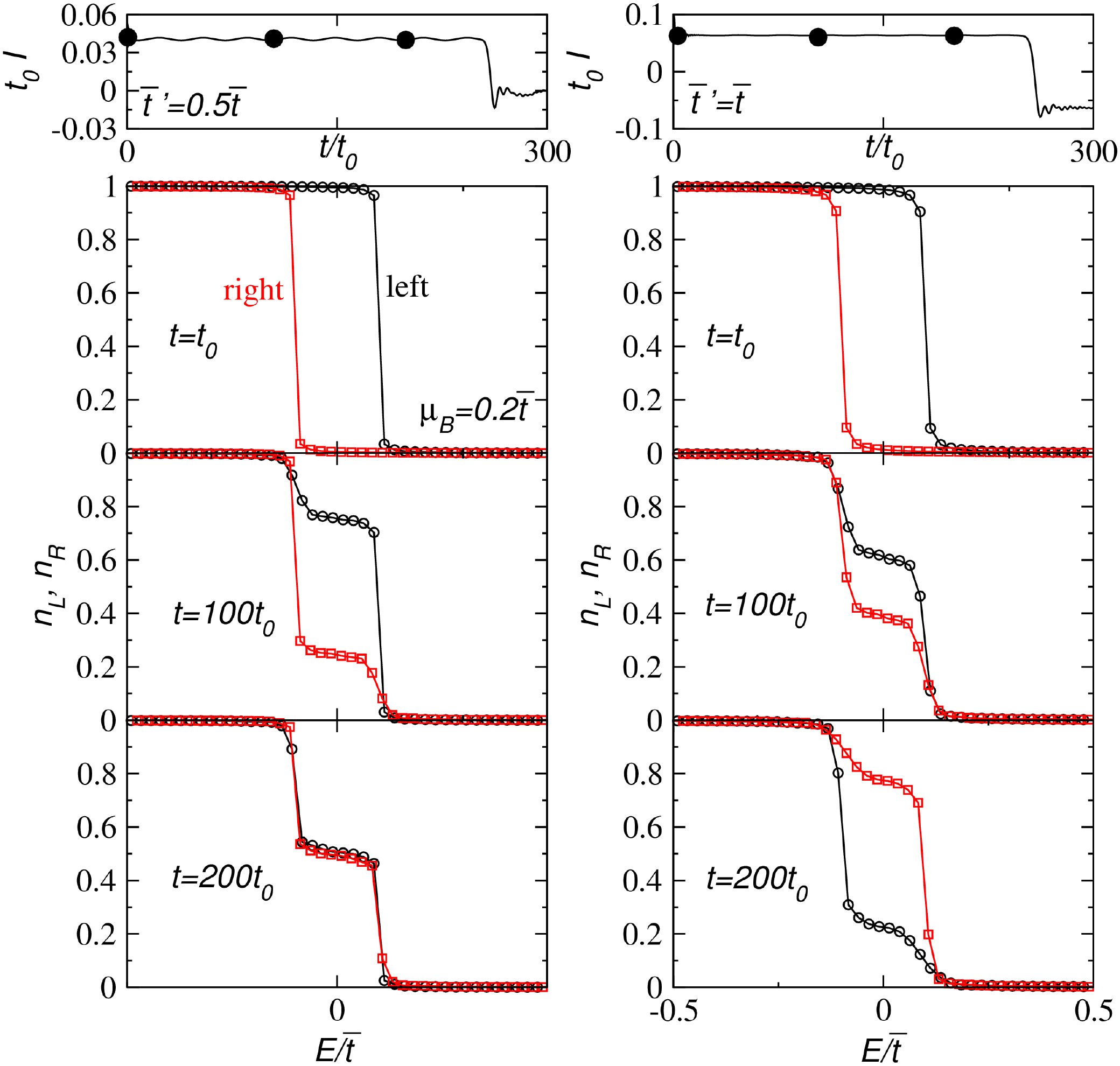} \caption{(Coor online) The distribution function of the weak-link case with
$\mu_{B}=0.2\bar{t}$ and $\bar{t}^{\prime}=0.5\bar{t}$ (left column)
and $\bar{t}$ (right column). From top to bottom, $t=t_{0}$, $100t_{0}$,
and $200t_{0}$. Here, $N=512$ (with the lattice initially half-filled) and the quasi-steady state current
persists to $240t_{0}$.}
\label{fig:dist_c5}
\end{figure}

\begin{figure}
\includegraphics[clip,width=3.4in]{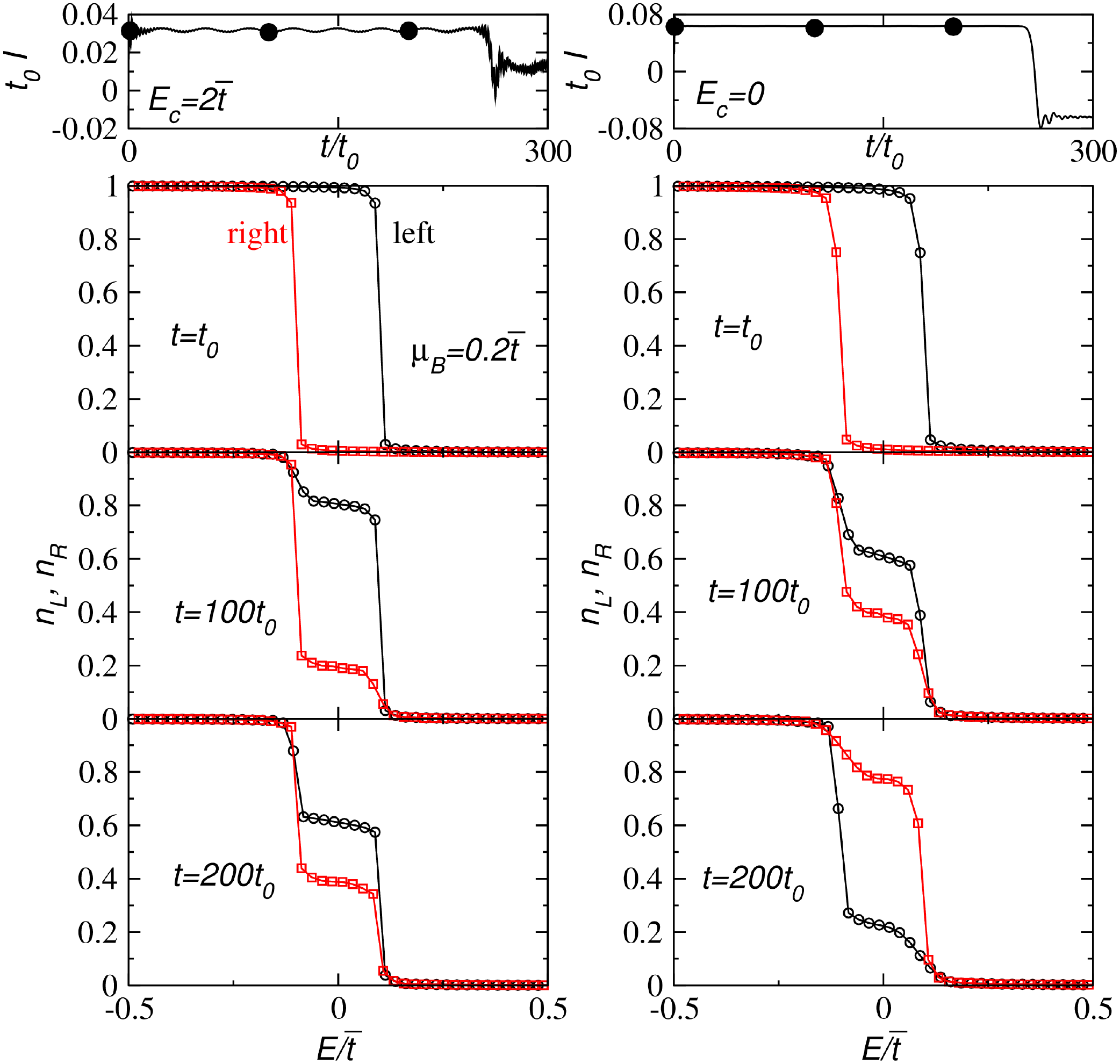} \caption{(Color online) The distribution function of the central-site case
with $\mu_{B}=0.2\bar{t}$ and $E_{c}=2\bar{t}$ (left column) and $0$
(right column). From top to bottom, $t=t_{0}$, $100t_{0}$, and $200t_{0}$.
Here, $N=512$ and the quasi-steady state current persists to $t=240t_{0}$.}
\label{fig:dist_c7}
\end{figure}

So far the MCF agree reasonably with Landauer formalism and FCS.
Now we will show several interesting phenomena associated with the finite size and conservation laws of isolated systems such as cold atoms.
We first study the particle distribution functions on the left
and the right sides. This can be done by first projecting the correlation matrix to the
left (right) half uniform lattice and obtaining $M_{L}=P_{L}CP_{L}$ and
$M_{R}=P_{R}CP_{R}$. Next we find the eigenvalues and the corresponding
unitary transformations of $H_{L}$ and $H_{R}$ (with the biases on) so
that $H_{L/R}=U_{L/R}D_{L/R}U_{L/R}^{\dagger}$, where $D_{L,R}=diag(\epsilon_{L/R,1},\cdots,\epsilon_{L/R,N/2})$.
Then we construct the correlation matrix in energy space and get
$\tilde{D}_{qq^{\prime}}^{L/R}=\sum_{i,j\in L/R}(U_{L/R}^{\dagger})_{qi}(U_{L/R})_{jq^{\prime}}(M_{L/R})_{ij}$.
For each eigenvalue $\epsilon_{L/R,q}$, the occupation number is
given by $n_{L/R}(\epsilon_{L/R,q})=\tilde{D}_{qq}^{L/R}$. In Fig.~\ref{fig:dist_c5}
we show the particle distributions for the weak-link case with $\bar{t}^{\prime}=0.5\bar{t}$
and $\bar{t}^{\prime}=\bar{t}$ at $t=t_{0}$, $100t_{0}$, and
$200t_{0}$ for $N=512$ with the system initially half-filled and $\mu_{B}=0.2\bar{t}$. The particle distributions
for the central-site case with similar parameters are shown in Fig.~\ref{fig:dist_c7}.

Clearly, the particle distributions on both sides vary dynamically but they
evolve in a coordinated fashion so that the current across the junction
remains constant for a long period of time. This is different from
the picture behind the Landauer formula. In Landauer formula the distributions
on the left (right) half lattice are fixed at $f_{L}$ ($f_{R}$) and
a constant tunneling constant $T$ determines the rate at which particles
move across the junction. On the other hand, for a finite system, the particle distributions must
evolve with time. 
If Eq.~\eqref{eq:Ianalytic} is naively used in this case, one may expect that the current decays with time because the difference between the distributions, $f_L(E)-f_R(E)$, should be a decreasing function when particles are flowing from the left to the right. In contrast, a plateau in the current emerges in the full quantum dynamics. 
Even more surprisingly, there exists a time interval when a QSSC still flows from left to right, yet the right lattice has more particles, as shown in the
bottom right panels of both Figs.~\ref{fig:dist_c5} and \ref{fig:dist_c7}
(for $t=200t_{0}$) \cite{population_note}. This highlights that this a highly correlated state that allows the QSSC to persist. We will see in the next
section this is due to causality and the finite speed of propagation of information, and thus is analogous to the light cone in special relativity.

There are recent proposals for designing batteries for atomtronic devices \cite{Zozulya}.
However, an important message from our study is that an isolated quantum system can
maintain a quasi steady-state current in many cases. The quasi-steady state, as we demonstrated, is
maintained by internal dynamics so a battery may not be the only way for generating a
steady current in atomtronic devices, one could instead engineer an appropriate initial state that will induce a QSSC.

\section{Light-cone of wave propagation}\label{sec:lightcone}
\begin{figure*}
\includegraphics[clip,width=16cm]{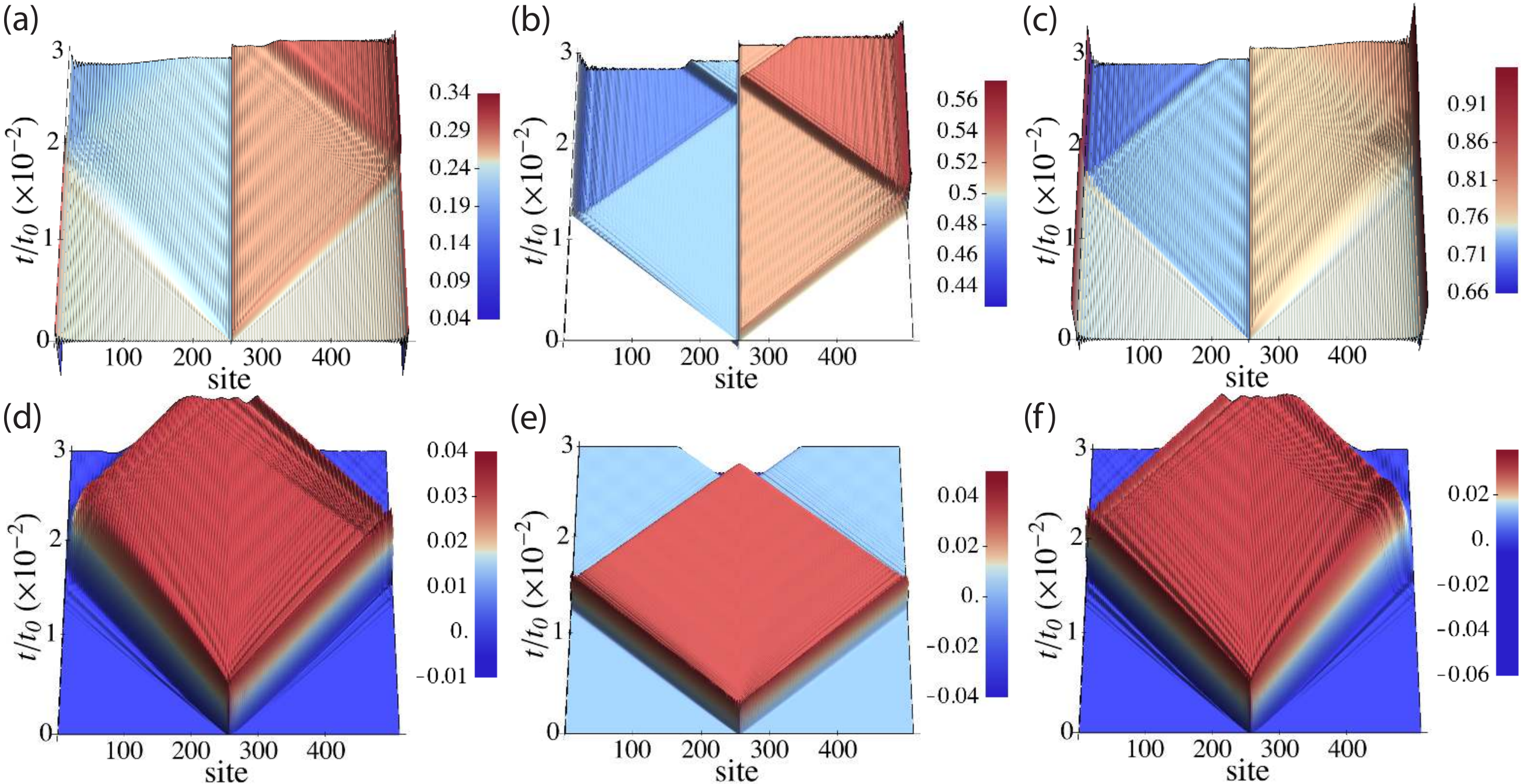}
\caption{(Color online) The density profiles (top row) and current profiles (bottom row) for a uniform chain driven out of equilibrium by a step-function bias $\mu_{B}=0.2\bar{t}$. Here $N=512$ with $N_p=128$ ((a) and (d)), $N_p=256$ ((b) and (e)), and $N_p=384$ ((c) and (f)). The Fermi velocity for half filling is $(2/t_0)$ and that for $(1/4)$ or $(3/4)$ filling is $(\sqrt{2}/t_0)$. \label{fig:c5_contour}}
\end{figure*}
\begin{figure*}
\includegraphics[clip,width=16cm]{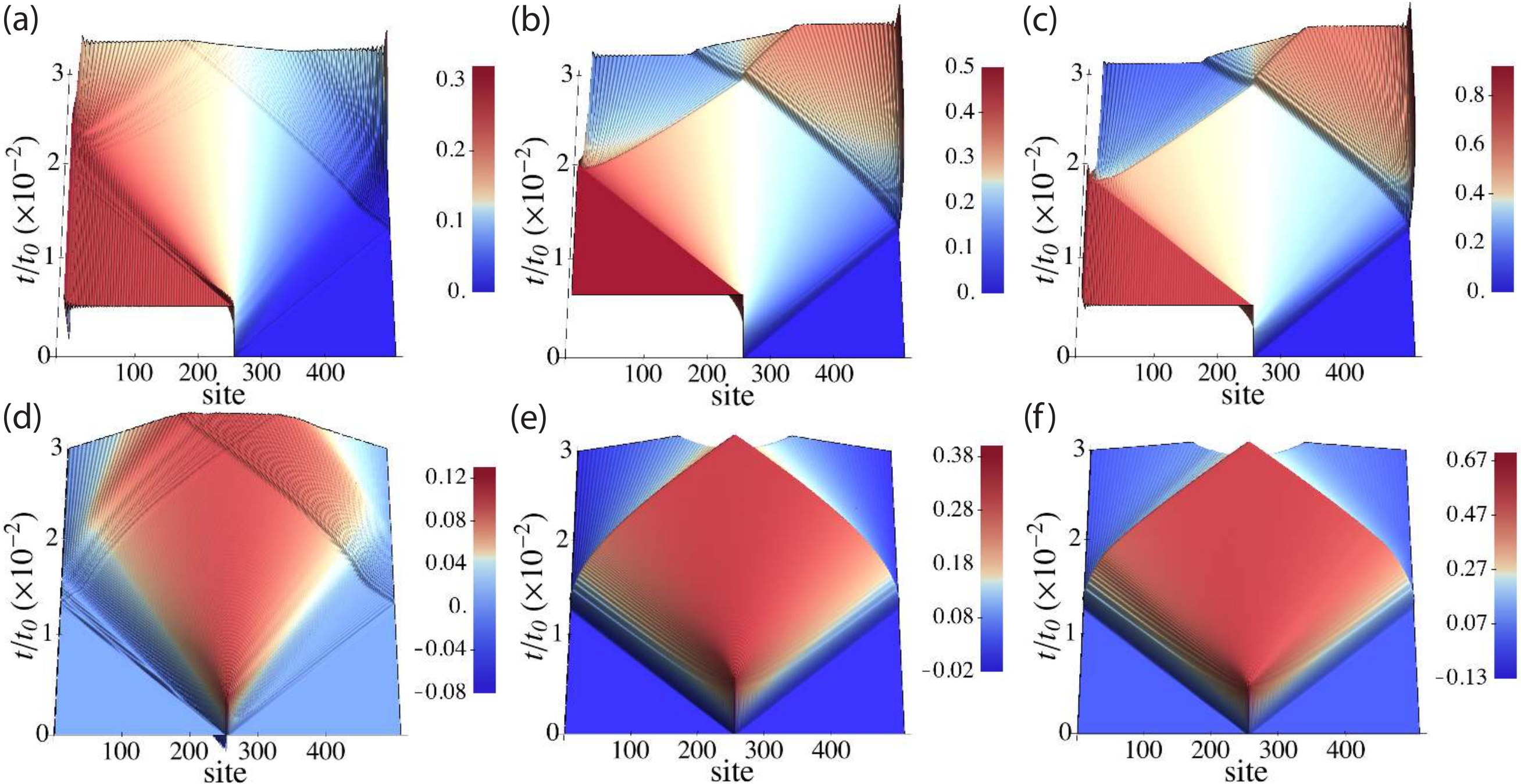}
\caption{(Color online) The density profiles (top row) and current profiles (bottom row) for a uniform chain driven out of equilibrium by suddenly blowing away particles on the right half. Here $N=512$ with initial particle number $N_p=128$ for (a) and (d), $N_p=256$ for (b) and (e), and $N_p=384$ for (c) and (f). \label{fig:c5_half_contour}}
\end{figure*}
In the last section, we saw that a QSSC can continue to flow even when the particle imbalance would indicate otherwise.
This effect is due to the finite speed of information. Recent experimental studies~\cite{LC1,Blochtransport} have shown that the density profile exhibits a ``light cone'' as an atomic cloud expands, and there are ongoing theoretical studies to support this fact \cite{Geiger14}. We can see this effect within the MCF (one of the many advantages of this formalism). We monitor the real-time dynamics of the density and current profiles for noninteracting fermions in a uniform lattice driven out of equilibrium by (1) a step-function potential as shown in Fig.~\ref{fig:Schematic}, and (2) a sudden removal of atoms on the right-half lattice as discussed in Ref.~\cite{MCFshort}. The time evolution of the first case is shown in Figure~\ref{fig:c5_contour} and that of the second case is shown in Fig.~\ref{fig:c5_half_contour}. In both cases one can see clearly a ``light cone'' within which the motion of atoms are confined. The propagation speed is limited by the Fermi velocity, which for filling $f$ is $v_F=2\sin(f\pi)/t_0$. For $N=512$ at half filling ($N_p=256$), it takes about $128t_0$ for the wave front to reach the boundary and reflect back. Around $256t_0$ the two wave fronts propagating in the opposite directions meet again in the middle. That is when the current stops showing the quasi steady-state behavior. This applies to both cases, as shown in Fig.~\ref{fig:c5_contour}(c) and (d) and Fig.~\ref{fig:c5_half_contour}(c) and (d). This explains the paradoxical behavior of the QSSC flowing counter to the particle imbalance. This happens because the information regarding the population imbalance still has not been carried to the junction region where the current is being monitored.

For the quarter filling ($N_p=128$), if the wave front propagates at the speed of the corresponding Fermi velocity $\sqrt{2}/t_0$, it takes about $181t_0$ for the wave front to reach the boundary and the two wave fronts meet again at around $362t_0$. Although the main body of the wave propagates at this speed, there are "leaks" of the wave which propagate at speed higher than $\sqrt{2}/t_0$ but they are limited by the maximal Fermi velocity $2/t_0$, as shown in Figs.~\ref{fig:c5_contour} and \ref{fig:c5_half_contour}. This ``leaking'' behavior is more prominent for the case of a sudden removal of half of the particles at higher filling. As shown in Fig.~\ref{fig:c5_half_contour}(e) and (f), for initial $(3/4)$ filling there is significant fraction of the wave propagation at $2/t_0$. For the step-function bias case (Fig.~\ref{fig:c5_contour}(e) and (f)), the main wave propagates at $\sqrt{2}/t_0$ and again the leak propagates at higher speed (limited by $2/t_0$). We also found that adding a weak central link or a central site with different onsite energy only decreases the magnitude of the current, but the speed of wave-front propagation remains the same for the same initial filling.

\section{Kubo Formalism}\label{sec:Kubo}
In order to connect the microcanonical and Landauer
approaches, we apply leading order perturbation theory on finite systems
by way of the Kubo formula \cite{Kubo_book,Fetter_book,DiVentrabook}
\begin{equation}
\avg{A\left(t\right)}=\avg A_{0}-\im\int_{0}^{t}dt^{\p}\avg{\left[\hat{A}\left(t\right),\hat{H}^{\p}\left(t^{\p}\right)\right]}\label{eq:Kubo}
\end{equation}
for the observable $A$. Here, $\hat{O}=e^{\im H_{0}t}Oe^{-\im H_{0}t}$
indicates an operator in the interaction picture, $H^{\p}$ is the
perturbing Hamiltonian, and $\avg O_{0}$ indicates an average with
respect to the initial state. For all practical purposes, here we use the wavefunctions of
finite-size systems without taking the thermodynamic limit commonly employed in
solid-sate systems.
We will consider a one-dimensional lattice
set out of equilibrium by connecting two initially disconnected halves
with a weak link or by the application of a bias to an initially connected
system, as shown in Fig. \ref{fig:Schematic}.

\subsection{Connecting the $L$ and $R$ lattices}
The initial Hamiltonian is
\[
H_{0}=H_{L}+H_{R},
\]
where
\[
H_{L}=-\sum_{\avg{i,j}}\bar{t}c_{i}^{\dg}c_{j}+\mu_{L}\sum_{i}c_{i}^{\dg}c_{i}
\]
and
\[
H_{R}=-\sum_{\avg{i,j}}\bar{t}d_{i}^{\dg}d_{j}-\mu_{R}\sum_{i}d_{i}^{\dg}d_{i}.
\]
The left and the right lattices are both finite lattices of length
$N$ with non-periodic (``open'') boundary conditions. We consider
the ground state of $H_{L}$ and $H_{R}$ fixed at half filling --
the bias can be thought of as added simultaneously with the connection
of the two lattices. The diagonalization of the left lattice is performed
by $c_{j}=\sum_{k}U_{jk}a_{k}$ with $U_{jk}=\sqrt{2/\left(N+1\right)}\sin\left(jk\pi/\left(N+1\right)\right)$
and $k=1,\,\ldots,\, N$, yielding $H_{L}=\sum_{k}\epsilon_{k}^{L}a_{k}^{\dg}a_{k}$
and $\epsilon_{k}^{L}=-2\bar{t}\cos\left(k\pi/\left(N+1\right)\right)+\mu_{L}$.
Similarly for the right lattice, using $d_{j}=\sum_{k}U_{jk}b_{k}$
gives $H_{R}=\sum_{k}\epsilon_{k}^{R}b_{k}^{\dg}b_{k}$ and $\epsilon_{k}^{R}=-2\bar{t}\cos\left(k\pi/\left(N+1\right)\right)+\mu_{R}$.

At $t=0$, the lattices are connected by the perturbing Hamiltonian
\[
H^{\p}=g\bar{t}\left(c_{1}^{\dg}d_{1}+d_{1}^{\dg}c_{1}\right),
\]
where $g=\bar{t}^{\prime}/\bar{t}$. Note that the numbering of the sites in both
lattices starts from the interface sites. The current is the quantity
of interest, hence we will take
\[
A=g\bar{t}c_{1}^{\dg}d_{1},
\]
where $A$ gives the hopping between the two halves of the lattice,
i.e., $c_{1}$ acts on the interface site on the left lattice and
$d_{1}$ on the interface site of the right lattice. This will give
the current through $I\left(t\right)=-2\mbox{Im}\avg{A\left(t\right)}$.

The interaction picture operators are
\[
\hat{A}\left(t\right)=g\bar{t}^2\sum_{k,k^{\p}}U_{k1}^{\dg}U_{1k^{\p}}e^{\im\left(\epsilon_{k}^{L}-\epsilon_{k^{\p}}^{R}\right)t}a_{k}^{\dg}b_{k^{\p}}
\]
and
\[
\hat{H}^{\p}\left(t^{\p}\right)=\left(\hat{A}\left(t^{\p}\right)+\hat{A}^{\dg}\left(t^{\p}\right)\right).
\]
Putting these into Eq. \eqref{eq:Kubo} and using that $\avg A_{0}=0$
for two initially disconnected lattices, we obtain
\[
\avg{A\left(t\right)}=g^2\bar{t}\sum_{k,k^{\p}}\left|U_{k1}\right|^{2}\left|U_{1k^{\p}}\right|^{2}\frac{n_{k}-n_{k^{\p}}}{\epsilon_{k}^{L}-\epsilon_{k^{\p}}^{R}}\left(1-e^{\im\left(\epsilon_{k}^{L}-\epsilon_{k^{\p}}^{R}\right)t}\right).
\]
The current is then
\begin{align}
I\left(t\right) & =-2\mbox{Im}\avg{A\left(t\right)}\nonumber \\
 & =2g^{2}\bar{t}^{2}\sum_{k,k^{\p}}\left|U_{k1}\right|^{2}\left|U_{1k^{\p}}\right|^{2}\frac{n_{k}-n_{k^{\p}}}{\epsilon_{k}^{L}-\epsilon_{k^{\p}}^{R}}\sin\left[\left(\epsilon_{k}^{L}-\epsilon_{k^{\p}}^{R}\right)t\right].
\end{align}
When the $L$ and $R$ lattices are half filled, this double sum will be nonzero when either $k\le N/2,\, k^{\p}>N/2$
or $k>N/2,\, k^{\p}\le N/2$.

To simplify the expressions and show the correspondence with Landauer, we take the semi-infinite limit for the left and right lattices obtaining
\begin{align*}
I\left(t\right) & =\frac{8g^{2}\bar{t}^{2}}{\pi^{2}}\int_{0}^{\pi}dk\int_{0}^{\pi}dk^{\p}\sin^{2}k\sin^{2}k^{\p}\\
 & \times\frac{n_{k}-n_{k^{\p}}}{\epsilon_{k}-\epsilon_{k^{\p}}+\mu_B}\sin\left[t\left(\epsilon_{k}-\epsilon_{k^{\p}}+\mu_B\right)\right],
\end{align*}
where $\epsilon_{k}=-2\bar{t}\cos k$ and $\mu_B=\mu_{L}-\mu_{R}$. At this
point, we have made no assumption about the strength of the bias,
the filling, or the temperature. We will now restrict ourselves to
the case of half filling and zero temperature.

\begin{figure}
\begin{centering}
\includegraphics[width=3.4in,clip]{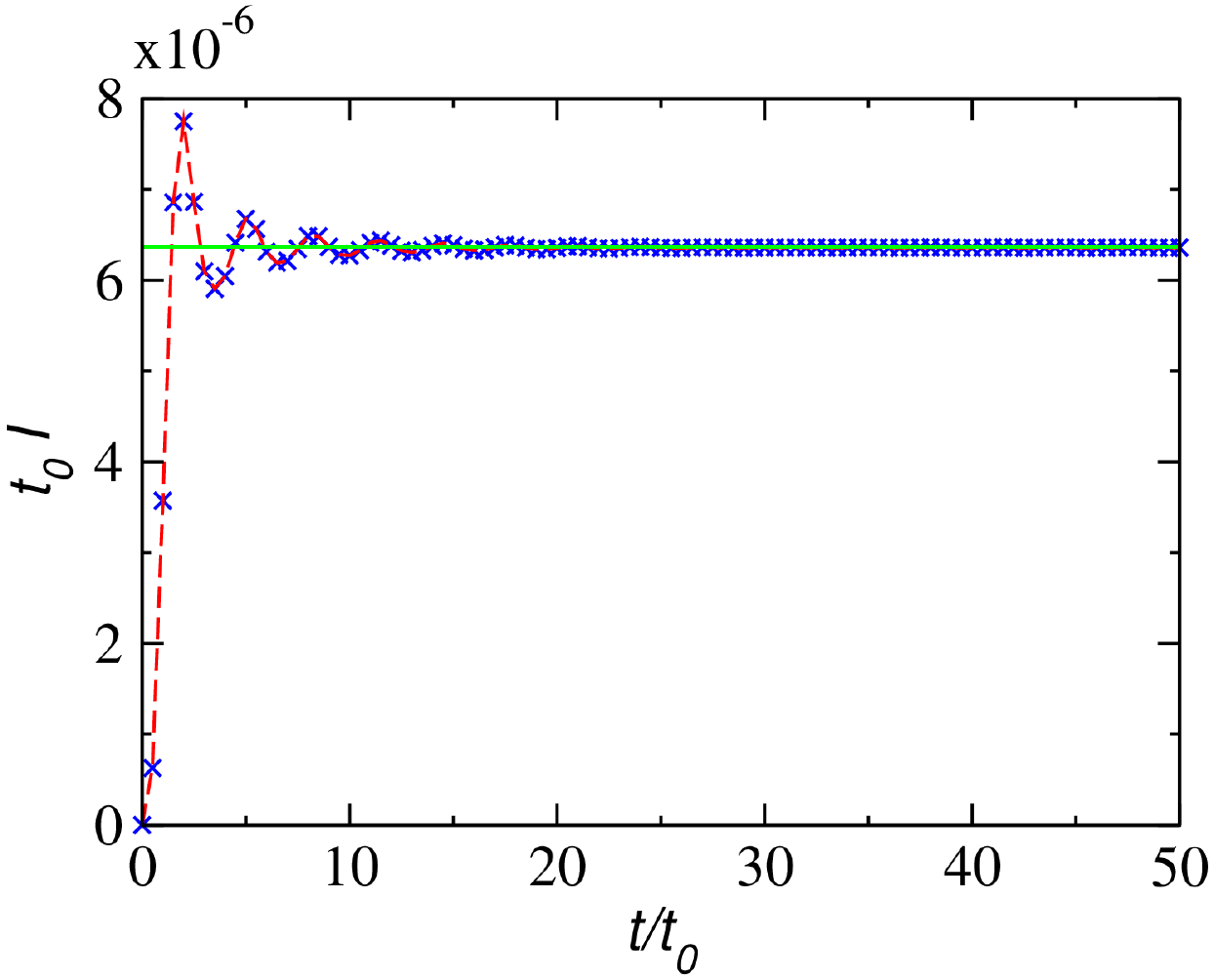}
\par\end{centering}
\caption{(Color online) Current versus time for connection-induced transport. The Kubo result
(blue crosses) compares very well with the exact microcanonical method
(red, dashed line), with both approaching the steady-state current
(green, dashed line) for long times. Here, the lattice
is of length 1600 sites, the bias is $\mu_B/\bar{t}=1/10$, and the strength
of the weak link is $g=1/100$. \label{fig:KuboWeakLink}}
\end{figure}

The two contributions to this expression give the forward and backward
currents, integrating over energy instead of wave vector,
\begin{align*}
I_{\rightarrow}\left(t\right)= & \frac{2g^{2}\bar{t}}{\pi^{2}}\int_{-2}^{0}d\epsilon\int_{0}^{2}d\epsilon^{\p}\left(1-\frac{\epsilon^{2}}{4}\right)^{1/2}\left(1-\frac{\epsilon^{\p2}}{4}\right)^{1/2}\\
 & \times\frac{\sin\left[t\left(\epsilon-\epsilon^{\p}+\mu_b\right)/t_0\right]}{\epsilon-\epsilon^{\p}+\mu_b}
\end{align*}
and
\begin{align*}
I_{\leftarrow}\left(t\right)= & \frac{2g^{2}\bar{t}}{\pi^{2}}\int_{0}^{2}d\epsilon\int_{-2}^{0}d\epsilon^{\p}\left(1-\frac{\epsilon^{2}}{4}\right)^{1/2}\left(1-\frac{\epsilon^{\p2}}{4}\right)^{1/2}\\
 & \times\frac{\sin\left[t\left(\epsilon-\epsilon^{\p}+\mu_b\right)/t_0\right]}{\epsilon-\epsilon^{\p}+\mu_b}.
\end{align*}
Here, $\mu_b=\mu_B/\bar{t}$. As $t\to\infty$, the fast oscillating function $\sin\left(tx\right)/\pi x$ enforces $\epsilon^{\p}=\epsilon+\mu_b$. This latter equality
can not be satisfied in the backward current as $\epsilon$ is positive
and $\mu_b$ is also positive, but $\epsilon^{\p}$ is negative. Thus,
only the forward current remains, giving
\[
I= \frac{g^{2}\bar{t}^{2}}{\pi}\int_{-\mu_b/2}^{\mu_b/2}d\epsilon \bar{g}_L \bar{g}_R
\]
in the steady state and including the factor of two for spin. Here, $\bar{g}_{L(R)}=\sqrt{4-\left(\epsilon\mp\mu_b/2\right)^{2}}$ and the bias is applied symmetrically. The result is insensitive, though, to how the bias is applied -- the left lattice can be shifted by $\mu_B$ and right lattice by $0$, or the left by $\mu_B/2$ and the right by $-\mu_B/2$. This expression is valid for arbitrary
bias and agrees with the Landauer expression, Eq.~\eqref{eq:LI_tp}, to leading order in $g$. For small bias, one obtains
\[
I\simeq\frac{4g^{2}\bar{t}}{\pi}\mu_{B}.
\]
Figure~\ref{fig:KuboWeakLink} shows the agreement of this expression with
the exact microcanonical expression for finite-size systems.

This Kubo approach is firmly rooted in the microcanonical picture -- we have a finite, closed system (the semi-infinite limit is taken only for convenience) set out of equilibrium. The resulting expressions separate out the short time behavior -- due to forward and backward fluctuations at all energy scales -- and the long time behavior -- the QSSC -- that emerges from just the forward current.

\subsection{Applied bias across the $L$ and $R$ lattices}
Let us now consider an initial Hamiltonian for a connected, homogeneous lattice of
length $2N$
\[
H_{0}=-\sum_{\avg{i,j}}\bar{t}c_{i}^{\dg}c_{j}.
\]
We consider the ground state of $H_{0}$ fixed at half filling. The
diagonalization is the same as above except with a lattice of length
$2N$, $c_{j}=\sum_{k}U_{jk}a_{k}$ with $U_{jk}=\sqrt{2/\left(2N+1\right)}\sin\left(jk\pi/\left(2N+1\right)\right)$
and $k=1,\,\ldots,\,2N$, yielding $H_{0}=\sum_{k}\epsilon_{k}a_{k}^{\dg}a_{k}$
and $\epsilon_{k}=-2\bar{t}\cos\left(k\pi/\left(2N+1\right)\right)$.

\begin{figure}
\begin{centering}
\includegraphics[width=3.4in,clip]{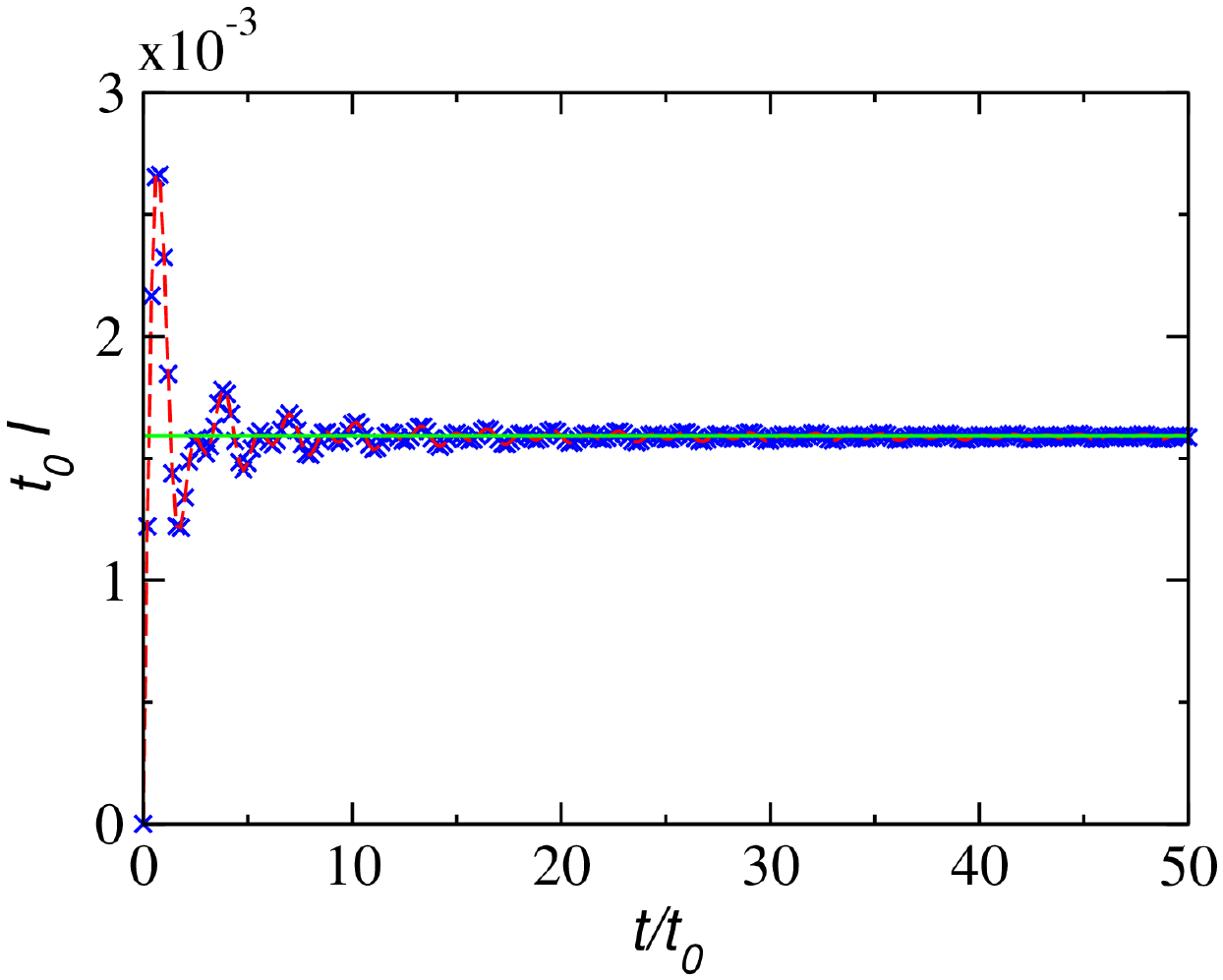}
\par\end{centering}
\caption{(Color online) Current versus time for bias-induced transport. The Kubo result (blue
crosses) compares very well with the exact microcanonical method (red,
dashed line), with both approaching the steady state current (green,
dashed line) for long times. Here, the lattice is of length 1600 sites,
the bias is $\mu_B/\bar{t}=1/100$. \label{fig:KuboBias}}
\end{figure}

At $t=0$, the lattices are perturbed by the Hamiltonian
\[
H^{\p}=\frac{\mu_B}{2}\sum_{i\in L}c_{i}^{\dg}c_{i}-\frac{\mu_B}{2}\sum_{i\in R}c_{i}^{\dg}c_{i},
\]
which applies the step potential bias as shown in Fig. \ref{fig:Schematic}.
The strength of the perturbation is the bias $\mu_B$. The current between
the two halves is of interest, and therefore we choose
\[
A=\bar{t}c_{N}^{\dg}c_{N+1}.
\]
The interaction picture operator is
\[
\hat{A}\left(t\right)=\sum_{k,k^{\p}}U_{kN}^{\dg}U_{N+1k^{\p}}e^{\im\left(\epsilon_{k}-\epsilon_{k^{\p}}\right)t}\bar{t}a_{k}^{\dg}a_{k^{\p}}.
\]
After some work, one finds that the current is
\begin{align*}
I\left(t\right) & =-2\mbox{Im}\avg{A\left(t\right)}\\
 & =\frac{2\mu_B\bar{t}}{\left(2N+1\right)^{2}}\sum_{\begin{array}{c} k\,\mathrm{Even}, \\ \, k^{\p}\,\mathrm{Odd}\end{array}}F_{kk^{\p}}\frac{n_{k}-n_{k^{\p}}}{\epsilon_{k}-\epsilon_{k^{\p}}}\sin\left[t\left(\epsilon_{k}-\epsilon_{k^{\p}}\right)\right],
\end{align*}
where
\[
F_{kk^{\p}}=2\bar{t}+\frac{4\bar{t}^2-\epsilon_{k}\epsilon_{k^{\p}}}{\epsilon_{k}-\epsilon_{k^{\p}}}.
\]
As $t\to\infty$, one can compute
the steady state current, $I \simeq \mu_B \bar{t}/\pi$, which includes a factor of two for spins. This agrees
with the Landauer expression, Eq.~\eqref{eq:LI_equal}, expanded to the leading order of $\mu_B$.
Figure \ref{fig:KuboBias}
shows the agreement of this expression with the exact microcanonical
expression for finite-size systems. One may build connections between the Landauer formalism and the MCF via the use of non-equilibrium Green's functions~\cite{Jauho94}. The Kubo approach here, however, gives explicit expressions for the effect of the reservoirs for discrete systems and thus explicitly connects closed, finite systems and their thermodynamic limit.

\section{Conclusion}\label{sec:conclusion}
In summary, we have discussed different theoretical viewpoints for quantum transport
phenomena that may be studied in ultra-cold atoms. In particular, we have compared the current, entanglement
entropy, and number fluctuations from the Landauer approach, semi-classical
FCS, and the micro-canonical formalism. In our study of two finite 1D lattices bridged by a junction, we
found a quasi steady-state current from the quantum dynamics in the
micro-canonical simulations. The magnitude of this quasi steady-state current agrees quantitatively with
the value predicted by the Landauer approach. The underlying mechanisms, nevertheless,
have been shown to be very different when the distributions of the two sides are analyzed. The distributions evolve in time and steadily deviate from Fermi-Dirac distributions even while the quasi steady-state current is maintained.

Our work points out several key issues when applying different formalisms
to closed quantum systems such as ultra-cold atoms in optical lattices. Of particular importance is the
confirmation, using the micro-canonical approach, that a quasi-steady state fermionic current can be established in a 1-D closed system for a finite period of time without the
need of inelastic effects or interaction effects beyond mean field~\cite{Bushong05}. The magnitude of this non-interacting
quasi-steady state current is independent of the way the bias is switched on. This also hints at the
fact that the Landauer formalism may not be the best suited for the study of transport properties of these finite closed systems, even though the average current that it predicts is correct for long times. This is because, in the case of elastic scattering, the current is dominated by local properties at the junction. On the other hand, other quantities of interest, such as the occupation of particles -- whether at a quasi-steady state or not --
are very sensitive to the full spatial extent of the wavefunctions.

The entanglement entropy from our simulations of the full quantum dynamics agrees with the formula
derived from semi-classical FCS with a binomial distribution, which raises the question of
how the wave nature of the transmitted particles can be well approximated
by such a distribution. We also found that quantum corrections are not significant
in the equal-time number fluctuations of these noninteracting systems. On the one hand, this supports the use of a semi-classical approach in
studying certain transport phenomena. On the other hand, finding transport coefficients that are
sensitive to quantum corrections is an interesting future direction.

Extending those
comparisons to higher dimensions or in the presence of interactions beyond mean field approximations could be very challenging, but they could lead to a deeper understanding of transport phenomena in closed quantum systems. We emphasize that these issues, such as the dynamics and feedback of reservoirs, quantum correlations, and matter-wave propagation, should be carefully investigated in more complex situations as we did here for non-interacting systems.

\begin{acknowledgments}
C.C.C. acknowledges the support of the U.S. Department of Energy through
the LANL/LDRD Program. MD acknowledges support from the DOE grant
DE-FG02-05ER46204.
\end{acknowledgments}

\bibliographystyle{apsrev4-1}

\begin{thebibliography}{50}%
\makeatletter
\providecommand \@ifxundefined [1]{%
 \@ifx{#1\undefined}
}%
\providecommand \@ifnum [1]{%
 \ifnum #1\expandafter \@firstoftwo
 \else \expandafter \@secondoftwo
 \fi
}%
\providecommand \@ifx [1]{%
 \ifx #1\expandafter \@firstoftwo
 \else \expandafter \@secondoftwo
 \fi
}%
\providecommand \natexlab [1]{#1}%
\providecommand \enquote  [1]{``#1''}%
\providecommand \bibnamefont  [1]{#1}%
\providecommand \bibfnamefont [1]{#1}%
\providecommand \citenamefont [1]{#1}%
\providecommand \href@noop [0]{\@secondoftwo}%
\providecommand \href [0]{\begingroup \@sanitize@url \@href}%
\providecommand \@href[1]{\@@startlink{#1}\@@href}%
\providecommand \@@href[1]{\endgroup#1\@@endlink}%
\providecommand \@sanitize@url [0]{\catcode `\\12\catcode `\$12\catcode
  `\&12\catcode `\#12\catcode `\^12\catcode `\_12\catcode `\%12\relax}%
\providecommand \@@startlink[1]{}%
\providecommand \@@endlink[0]{}%
\providecommand \url  [0]{\begingroup\@sanitize@url \@url }%
\providecommand \@url [1]{\endgroup\@href {#1}{\urlprefix }}%
\providecommand \urlprefix  [0]{URL }%
\providecommand \Eprint [0]{\href }%
\providecommand \doibase [0]{http://dx.doi.org/}%
\providecommand \selectlanguage [0]{\@gobble}%
\providecommand \bibinfo  [0]{\@secondoftwo}%
\providecommand \bibfield  [0]{\@secondoftwo}%
\providecommand \translation [1]{[#1]}%
\providecommand \BibitemOpen [0]{}%
\providecommand \bibitemStop [0]{}%
\providecommand \bibitemNoStop [0]{.\EOS\space}%
\providecommand \EOS [0]{\spacefactor3000\relax}%
\providecommand \BibitemShut  [1]{\csname bibitem#1\endcsname}%
\let\auto@bib@innerbib\@empty
\bibitem [{\citenamefont {Ott}\ \emph {et~al.}(2004)\citenamefont {Ott},
  \citenamefont {de~Mirandes}, \citenamefont {Ferlaino}, \citenamefont {Roati},
  \citenamefont {Modugno},\ and\ \citenamefont {Inguscio}}]{Ott04}%
  \BibitemOpen
  \bibfield  {author} {\bibinfo {author} {\bibfnamefont {H.}~\bibnamefont
  {Ott}}, \bibinfo {author} {\bibfnamefont {E.}~\bibnamefont {de~Mirandes}},
  \bibinfo {author} {\bibfnamefont {F.}~\bibnamefont {Ferlaino}}, \bibinfo
  {author} {\bibfnamefont {G.}~\bibnamefont {Roati}}, \bibinfo {author}
  {\bibfnamefont {G.}~\bibnamefont {Modugno}}, \ and\ \bibinfo {author}
  {\bibfnamefont {M.}~\bibnamefont {Inguscio}},\ }\href@noop {} {\bibfield
  {journal} {\bibinfo  {journal} {Phys. Rev. Lett.}\ }\textbf {\bibinfo
  {volume} {92}},\ \bibinfo {pages} {160601} (\bibinfo {year}
  {2004})}\BibitemShut {NoStop}%
\bibitem [{\citenamefont {Salger}\ \emph {et~al.}(2009)\citenamefont {Salger},
  \citenamefont {Kling}, \citenamefont {Hecking}, \citenamefont {Geckeler},
  \citenamefont {Morales-Molina},\ and\ \citenamefont {Weitz}}]{Salger09}%
  \BibitemOpen
  \bibfield  {author} {\bibinfo {author} {\bibfnamefont {T.}~\bibnamefont
  {Salger}}, \bibinfo {author} {\bibfnamefont {S.}~\bibnamefont {Kling}},
  \bibinfo {author} {\bibfnamefont {T.}~\bibnamefont {Hecking}}, \bibinfo
  {author} {\bibfnamefont {C.}~\bibnamefont {Geckeler}}, \bibinfo {author}
  {\bibfnamefont {L.}~\bibnamefont {Morales-Molina}}, \ and\ \bibinfo {author}
  {\bibfnamefont {M.}~\bibnamefont {Weitz}},\ }\href@noop {} {\bibfield
  {journal} {\bibinfo  {journal} {Science}\ }\textbf {\bibinfo {volume}
  {326}},\ \bibinfo {pages} {1241} (\bibinfo {year} {2009})}\BibitemShut
  {NoStop}%
\bibitem [{\citenamefont {Schneider}\ \emph {et~al.}(2012)\citenamefont
  {Schneider}, \citenamefont {Hackermuller}, \citenamefont {Ronzheimer},
  \citenamefont {Will}, \citenamefont {Braun}, \citenamefont {Best},
  \citenamefont {Bloch}, \citenamefont {Demler}, \citenamefont {Mandt},
  \citenamefont {Rasch},\ and\ \citenamefont {Rosch}}]{Blochtransport}%
  \BibitemOpen
  \bibfield  {author} {\bibinfo {author} {\bibfnamefont {U.}~\bibnamefont
  {Schneider}}, \bibinfo {author} {\bibfnamefont {L.}~\bibnamefont
  {Hackermuller}}, \bibinfo {author} {\bibfnamefont {J.~P.}\ \bibnamefont
  {Ronzheimer}}, \bibinfo {author} {\bibfnamefont {S.}~\bibnamefont {Will}},
  \bibinfo {author} {\bibfnamefont {S.}~\bibnamefont {Braun}}, \bibinfo
  {author} {\bibfnamefont {T.}~\bibnamefont {Best}}, \bibinfo {author}
  {\bibfnamefont {I.}~\bibnamefont {Bloch}}, \bibinfo {author} {\bibfnamefont
  {E.}~\bibnamefont {Demler}}, \bibinfo {author} {\bibfnamefont
  {S.}~\bibnamefont {Mandt}}, \bibinfo {author} {\bibfnamefont
  {D.}~\bibnamefont {Rasch}}, \ and\ \bibinfo {author} {\bibfnamefont
  {A.}~\bibnamefont {Rosch}},\ }\href@noop {} {\bibfield  {journal} {\bibinfo
  {journal} {Nat. Phys.}\ }\textbf {\bibinfo {volume} {8}},\ \bibinfo {pages}
  {213} (\bibinfo {year} {2012})}\BibitemShut {NoStop}%
\bibitem [{\citenamefont {Pepino}\ \emph {et~al.}(2009)\citenamefont {Pepino},
  \citenamefont {Cooper}, \citenamefont {Anderson},\ and\ \citenamefont
  {Holland}}]{Atomtronics1}%
  \BibitemOpen
  \bibfield  {author} {\bibinfo {author} {\bibfnamefont {R.~A.}\ \bibnamefont
  {Pepino}}, \bibinfo {author} {\bibfnamefont {J.}~\bibnamefont {Cooper}},
  \bibinfo {author} {\bibfnamefont {D.~Z.}\ \bibnamefont {Anderson}}, \ and\
  \bibinfo {author} {\bibfnamefont {M.~J.}\ \bibnamefont {Holland}},\
  }\href@noop {} {\bibfield  {journal} {\bibinfo  {journal} {Phys. Rev. Lett.}\
  }\textbf {\bibinfo {volume} {103}},\ \bibinfo {pages} {140405} (\bibinfo
  {year} {2009})}\BibitemShut {NoStop}%
\bibitem [{\citenamefont {Ramanathan}\ \emph {et~al.}(2011)\citenamefont
  {Ramanathan}, \citenamefont {Wright}, \citenamefont {Muniz}, \citenamefont
  {Zelan}, \citenamefont {Hill~III}, \citenamefont {Lobb}, \citenamefont
  {Helmerson}, \citenamefont {Phillips},\ and\ \citenamefont
  {Campbell}}]{ring1}%
  \BibitemOpen
  \bibfield  {author} {\bibinfo {author} {\bibfnamefont {A.}~\bibnamefont
  {Ramanathan}}, \bibinfo {author} {\bibfnamefont {K.~C.}\ \bibnamefont
  {Wright}}, \bibinfo {author} {\bibfnamefont {S.~R.}\ \bibnamefont {Muniz}},
  \bibinfo {author} {\bibfnamefont {M.}~\bibnamefont {Zelan}}, \bibinfo
  {author} {\bibfnamefont {W.~T.}\ \bibnamefont {Hill~III}}, \bibinfo {author}
  {\bibfnamefont {C.~J.}\ \bibnamefont {Lobb}}, \bibinfo {author}
  {\bibfnamefont {K.}~\bibnamefont {Helmerson}}, \bibinfo {author}
  {\bibfnamefont {W.~D.}\ \bibnamefont {Phillips}}, \ and\ \bibinfo {author}
  {\bibfnamefont {G.~K.}\ \bibnamefont {Campbell}},\ }\href@noop {} {\bibfield
  {journal} {\bibinfo  {journal} {Phys. Rev. Lett.}\ }\textbf {\bibinfo
  {volume} {106}},\ \bibinfo {pages} {130401} (\bibinfo {year}
  {2011})}\BibitemShut {NoStop}%
\bibitem [{\citenamefont {Brantut}\ \emph {et~al.}(2012)\citenamefont
  {Brantut}, \citenamefont {Meineke}, \citenamefont {Stadler}, \citenamefont
  {Krinner},\ and\ \citenamefont {Esslinger}}]{Esslinger12}%
  \BibitemOpen
  \bibfield  {author} {\bibinfo {author} {\bibfnamefont {J.~P.}\ \bibnamefont
  {Brantut}}, \bibinfo {author} {\bibfnamefont {J.}~\bibnamefont {Meineke}},
  \bibinfo {author} {\bibfnamefont {D.}~\bibnamefont {Stadler}}, \bibinfo
  {author} {\bibfnamefont {S.}~\bibnamefont {Krinner}}, \ and\ \bibinfo
  {author} {\bibfnamefont {T.}~\bibnamefont {Esslinger}},\ }\href@noop {}
  {\bibfield  {journal} {\bibinfo  {journal} {Science}\ }\textbf {\bibinfo
  {volume} {337}},\ \bibinfo {pages} {1069} (\bibinfo {year}
  {2012})}\BibitemShut {NoStop}%
\bibitem [{\citenamefont {Lee}\ \emph {et~al.}(2013)\citenamefont {Lee},
  \citenamefont {Mcllvain}, \citenamefont {Lobb},\ and\ \citenamefont
  {Hill}}]{Hill13}%
  \BibitemOpen
  \bibfield  {author} {\bibinfo {author} {\bibfnamefont {J.~G.}\ \bibnamefont
  {Lee}}, \bibinfo {author} {\bibfnamefont {B.~J.}\ \bibnamefont {Mcllvain}},
  \bibinfo {author} {\bibfnamefont {C.~J.}\ \bibnamefont {Lobb}}, \ and\
  \bibinfo {author} {\bibfnamefont {W.~T.}\ \bibnamefont {Hill}},\ }\href@noop
  {} {\bibfield  {journal} {\bibinfo  {journal} {Sci. Rep.}\ }\textbf {\bibinfo
  {volume} {3}},\ \bibinfo {pages} {1034} (\bibinfo {year} {2013})}\BibitemShut
  {NoStop}%
\bibitem [{\citenamefont {Zozulya}\ and\ \citenamefont
  {Anderson}(2013)}]{Zozulya}%
  \BibitemOpen
  \bibfield  {author} {\bibinfo {author} {\bibfnamefont {A.~A.}\ \bibnamefont
  {Zozulya}}\ and\ \bibinfo {author} {\bibfnamefont {D.~Z.}\ \bibnamefont
  {Anderson}},\ }\href@noop {} {\bibfield  {journal} {\bibinfo  {journal}
  {Phys. Rev. A}\ }\textbf {\bibinfo {volume} {88}},\ \bibinfo {pages} {043641}
  (\bibinfo {year} {2013})}\BibitemShut {NoStop}%
\bibitem [{\citenamefont {Eckel}\ \emph {et~al.}(2014)\citenamefont {Eckel},
  \citenamefont {Lee}, \citenamefont {Jendrzejewski}, \citenamefont {Murray},
  \citenamefont {Clark}, \citenamefont {Lobb}, \citenamefont {Phillips},
  \citenamefont {Edwards},\ and\ \citenamefont {Campbell}}]{Eckel14}%
  \BibitemOpen
  \bibfield  {author} {\bibinfo {author} {\bibfnamefont {S.}~\bibnamefont
  {Eckel}}, \bibinfo {author} {\bibfnamefont {J.~G.}\ \bibnamefont {Lee}},
  \bibinfo {author} {\bibfnamefont {F.}~\bibnamefont {Jendrzejewski}}, \bibinfo
  {author} {\bibfnamefont {N.}~\bibnamefont {Murray}}, \bibinfo {author}
  {\bibfnamefont {C.~W.}\ \bibnamefont {Clark}}, \bibinfo {author}
  {\bibfnamefont {C.~J.}\ \bibnamefont {Lobb}}, \bibinfo {author}
  {\bibfnamefont {W.~D.}\ \bibnamefont {Phillips}}, \bibinfo {author}
  {\bibfnamefont {M.}~\bibnamefont {Edwards}}, \ and\ \bibinfo {author}
  {\bibfnamefont {G.~K.}\ \bibnamefont {Campbell}},\ }\href@noop {} {\bibfield
  {journal} {\bibinfo  {journal} {Nature}\ }\textbf {\bibinfo {volume} {506}},\
  \bibinfo {pages} {200} (\bibinfo {year} {2014})}\BibitemShut {NoStop}%
\bibitem [{\citenamefont {Landauer}(1957)}]{Landauer}%
  \BibitemOpen
  \bibfield  {author} {\bibinfo {author} {\bibfnamefont {R.}~\bibnamefont
  {Landauer}},\ }\href@noop {} {\bibfield  {journal} {\bibinfo  {journal} {IBM
  J. Res. Dev.}\ }\textbf {\bibinfo {volume} {1}},\ \bibinfo {pages} {223}
  (\bibinfo {year} {1957})}\BibitemShut {NoStop}%
\bibitem [{\citenamefont {Ventra}(2008)}]{DiVentrabook}%
  \BibitemOpen
  \bibfield  {author} {\bibinfo {author} {\bibfnamefont {M.~D.}\ \bibnamefont
  {Ventra}},\ }\href@noop {} {\emph {\bibinfo {title} {Electrical Transport in
  nanoscale systems}}}\ (\bibinfo  {publisher} {Cambridge University Press},\
  \bibinfo {address} {Cambridge, UK},\ \bibinfo {year} {2008})\BibitemShut
  {NoStop}%
\bibitem [{\citenamefont {Bruun}\ and\ \citenamefont {Smith}(2005)}]{Bruun05}%
  \BibitemOpen
  \bibfield  {author} {\bibinfo {author} {\bibfnamefont {G.~M.}\ \bibnamefont
  {Bruun}}\ and\ \bibinfo {author} {\bibfnamefont {H.}~\bibnamefont {Smith}},\
  }\href@noop {} {\bibfield  {journal} {\bibinfo  {journal} {Phys. Rev. A}\
  }\textbf {\bibinfo {volume} {72}},\ \bibinfo {pages} {043605} (\bibinfo
  {year} {2005})}\BibitemShut {NoStop}%
\bibitem [{\citenamefont {Schafer}\ and\ \citenamefont
  {Teaney}(2009)}]{SchaferRoPP}%
  \BibitemOpen
  \bibfield  {author} {\bibinfo {author} {\bibfnamefont {T.}~\bibnamefont
  {Schafer}}\ and\ \bibinfo {author} {\bibfnamefont {D.}~\bibnamefont
  {Teaney}},\ }\href@noop {} {\bibfield  {journal} {\bibinfo  {journal} {Rep.
  Prog. Phys.}\ }\textbf {\bibinfo {volume} {72}},\ \bibinfo {pages} {126001}
  (\bibinfo {year} {2009})}\BibitemShut {NoStop}%
\bibitem [{\citenamefont {Brantut}\ \emph {et~al.}(2013)\citenamefont
  {Brantut}, \citenamefont {Grenier}, \citenamefont {Meineke}, \citenamefont
  {Stadler}, \citenamefont {Krinner}, \citenamefont {Kollath}, \citenamefont
  {Esslinger},\ and\ \citenamefont {Georges}}]{Brantut}%
  \BibitemOpen
  \bibfield  {author} {\bibinfo {author} {\bibfnamefont {J.~P.}\ \bibnamefont
  {Brantut}}, \bibinfo {author} {\bibfnamefont {C.}~\bibnamefont {Grenier}},
  \bibinfo {author} {\bibfnamefont {J.}~\bibnamefont {Meineke}}, \bibinfo
  {author} {\bibfnamefont {D.}~\bibnamefont {Stadler}}, \bibinfo {author}
  {\bibfnamefont {S.}~\bibnamefont {Krinner}}, \bibinfo {author} {\bibfnamefont
  {C.}~\bibnamefont {Kollath}}, \bibinfo {author} {\bibfnamefont
  {T.}~\bibnamefont {Esslinger}}, \ and\ \bibinfo {author} {\bibfnamefont
  {A.}~\bibnamefont {Georges}},\ }\href@noop {} {\bibfield  {journal} {\bibinfo
   {journal} {Science}\ }\textbf {\bibinfo {volume} {342}},\ \bibinfo {pages}
  {713} (\bibinfo {year} {2013})}\BibitemShut {NoStop}%
\bibitem [{\citenamefont {Bruderer}\ and\ \citenamefont
  {Belzig}(2012)}]{Bruderer12}%
  \BibitemOpen
  \bibfield  {author} {\bibinfo {author} {\bibfnamefont {M.}~\bibnamefont
  {Bruderer}}\ and\ \bibinfo {author} {\bibfnamefont {W.}~\bibnamefont
  {Belzig}},\ }\href@noop {} {\bibfield  {journal} {\bibinfo  {journal} {Phys.
  Rev. A}\ }\textbf {\bibinfo {volume} {85}},\ \bibinfo {pages} {013623}
  (\bibinfo {year} {2012})}\BibitemShut {NoStop}%
\bibitem [{\citenamefont {Gutman}\ \emph {et~al.}(2012)\citenamefont {Gutman},
  \citenamefont {Gefen},\ and\ \citenamefont {Mirlin}}]{Gutman12}%
  \BibitemOpen
  \bibfield  {author} {\bibinfo {author} {\bibfnamefont {D.~B.}\ \bibnamefont
  {Gutman}}, \bibinfo {author} {\bibfnamefont {Y.}~\bibnamefont {Gefen}}, \
  and\ \bibinfo {author} {\bibfnamefont {A.~D.}\ \bibnamefont {Mirlin}},\
  }\href@noop {} {\bibfield  {journal} {\bibinfo  {journal} {Phys. Rev. B}\
  }\textbf {\bibinfo {volume} {85}},\ \bibinfo {pages} {125102} (\bibinfo
  {year} {2012})}\BibitemShut {NoStop}%
\bibitem [{\citenamefont {Simpson}\ \emph {et~al.}(2014)\citenamefont
  {Simpson}, \citenamefont {Gangardt}, \citenamefont {Lerner},\ and\
  \citenamefont {Kruger}}]{Simpson14}%
  \BibitemOpen
  \bibfield  {author} {\bibinfo {author} {\bibfnamefont {D.~P.}\ \bibnamefont
  {Simpson}}, \bibinfo {author} {\bibfnamefont {D.~M.}\ \bibnamefont
  {Gangardt}}, \bibinfo {author} {\bibfnamefont {I.~V.}\ \bibnamefont
  {Lerner}}, \ and\ \bibinfo {author} {\bibfnamefont {P.}~\bibnamefont
  {Kruger}},\ }\href@noop {} {\bibfield  {journal} {\bibinfo  {journal} {Phys.
  Rev. Lett.}\ }\textbf {\bibinfo {volume} {112}},\ \bibinfo {pages} {100601}
  (\bibinfo {year} {2014})}\BibitemShut {NoStop}%
\bibitem [{Kli()}]{KlichLevitov}%
  \BibitemOpen
  \href@noop {} {}\bibinfo {note} {I. Klich and L. Levitov, Phys. Rev. Lett.
  \textbf{102}, 100502 (2009); Advances in Theoretical Physics: Landau Memorial
  Conference, Edited by V. Lebedev and M. V. Feigelman (AIP Conference
  Proceedings), volume 1134, 36-45 (2009)}\BibitemShut {NoStop}%
\bibitem [{\citenamefont {Di~Ventra}\ and\ \citenamefont
  {Todorov}(2005)}]{micro}%
  \BibitemOpen
  \bibfield  {author} {\bibinfo {author} {\bibfnamefont {M.}~\bibnamefont
  {Di~Ventra}}\ and\ \bibinfo {author} {\bibfnamefont {T.}~\bibnamefont
  {Todorov}},\ }\href@noop {} {\bibfield  {journal} {\bibinfo  {journal} {J.
  Phys. Cond. Matt.}\ }\textbf {\bibinfo {volume} {16}},\ \bibinfo {pages}
  {8025} (\bibinfo {year} {2005})}\BibitemShut {NoStop}%
\bibitem [{\citenamefont {Bushong}\ \emph {et~al.}(2005)\citenamefont
  {Bushong}, \citenamefont {Sai},\ and\ \citenamefont {Di~Ventra}}]{Bushong05}%
  \BibitemOpen
  \bibfield  {author} {\bibinfo {author} {\bibfnamefont {N.}~\bibnamefont
  {Bushong}}, \bibinfo {author} {\bibfnamefont {N.}~\bibnamefont {Sai}}, \ and\
  \bibinfo {author} {\bibfnamefont {M.}~\bibnamefont {Di~Ventra}},\ }\href@noop
  {} {\bibfield  {journal} {\bibinfo  {journal} {Nano Lett.}\ }\textbf
  {\bibinfo {volume} {5}},\ \bibinfo {pages} {2569} (\bibinfo {year}
  {2005})}\BibitemShut {NoStop}%
\bibitem [{\citenamefont {Kurth}\ \emph {et~al.}(2005)\citenamefont {Kurth},
  \citenamefont {Stefanucci}, \citenamefont {Almbladh}, \citenamefont {Rubio},\
  and\ \citenamefont {Gross}}]{Gross05}%
  \BibitemOpen
  \bibfield  {author} {\bibinfo {author} {\bibfnamefont {S.}~\bibnamefont
  {Kurth}}, \bibinfo {author} {\bibfnamefont {G.}~\bibnamefont {Stefanucci}},
  \bibinfo {author} {\bibfnamefont {C.~O.}\ \bibnamefont {Almbladh}}, \bibinfo
  {author} {\bibfnamefont {A.}~\bibnamefont {Rubio}}, \ and\ \bibinfo {author}
  {\bibfnamefont {E.~K.~U.}\ \bibnamefont {Gross}},\ }\href@noop {} {\bibfield
  {journal} {\bibinfo  {journal} {Phys. Rev. B}\ }\textbf {\bibinfo {volume}
  {72}},\ \bibinfo {pages} {035308} (\bibinfo {year} {2005})}\BibitemShut
  {NoStop}%
\bibitem [{\citenamefont {Cheng}\ \emph {et~al.}(2006)\citenamefont {Cheng},
  \citenamefont {Evans},\ and\ \citenamefont {Voorhis}}]{VanVoorhis06}%
  \BibitemOpen
  \bibfield  {author} {\bibinfo {author} {\bibfnamefont {C.~L.}\ \bibnamefont
  {Cheng}}, \bibinfo {author} {\bibfnamefont {J.~S.}\ \bibnamefont {Evans}}, \
  and\ \bibinfo {author} {\bibfnamefont {T.~V.}\ \bibnamefont {Voorhis}},\
  }\href@noop {} {\bibfield  {journal} {\bibinfo  {journal} {Phys. Rev. B}\
  }\textbf {\bibinfo {volume} {74}},\ \bibinfo {pages} {155112} (\bibinfo
  {year} {2006})}\BibitemShut {NoStop}%
\bibitem [{\citenamefont {Chien}\ \emph {et~al.}(2012)\citenamefont {Chien},
  \citenamefont {Zwolak},\ and\ \citenamefont {Di~Ventra}}]{MCFshort}%
  \BibitemOpen
  \bibfield  {author} {\bibinfo {author} {\bibfnamefont {C.~C.}\ \bibnamefont
  {Chien}}, \bibinfo {author} {\bibfnamefont {M.}~\bibnamefont {Zwolak}}, \
  and\ \bibinfo {author} {\bibfnamefont {M.}~\bibnamefont {Di~Ventra}},\
  }\href@noop {} {\bibfield  {journal} {\bibinfo  {journal} {Phys. Rev. A}\
  }\textbf {\bibinfo {volume} {85}},\ \bibinfo {pages} {041601(R)} (\bibinfo
  {year} {2012})}\BibitemShut {NoStop}%
\bibitem [{\citenamefont {Chien}\ and\ \citenamefont
  {Di~Ventra}(2012)}]{MCF_TD}%
  \BibitemOpen
  \bibfield  {author} {\bibinfo {author} {\bibfnamefont {C.~C.}\ \bibnamefont
  {Chien}}\ and\ \bibinfo {author} {\bibfnamefont {M.}~\bibnamefont
  {Di~Ventra}},\ }\href@noop {} {\bibfield  {journal} {\bibinfo  {journal}
  {EPL}\ }\textbf {\bibinfo {volume} {99}},\ \bibinfo {pages} {40003} (\bibinfo
  {year} {2012})}\BibitemShut {NoStop}%
\bibitem [{\citenamefont {Chien}\ and\ \citenamefont
  {Di~Ventra}(2013)}]{PhaseTran13}%
  \BibitemOpen
  \bibfield  {author} {\bibinfo {author} {\bibfnamefont {C.~C.}\ \bibnamefont
  {Chien}}\ and\ \bibinfo {author} {\bibfnamefont {M.}~\bibnamefont
  {Di~Ventra}},\ }\href@noop {} {\bibfield  {journal} {\bibinfo  {journal}
  {Phys. Rev. A}\ }\textbf {\bibinfo {volume} {87}},\ \bibinfo {pages} {023609}
  (\bibinfo {year} {2013})}\BibitemShut {NoStop}%
\bibitem [{\citenamefont {Chien}\ \emph {et~al.}(2013)\citenamefont {Chien},
  \citenamefont {Gruss}, \citenamefont {Di~Ventra},\ and\ \citenamefont
  {Zwolak}}]{int_induced}%
  \BibitemOpen
  \bibfield  {author} {\bibinfo {author} {\bibfnamefont {C.~C.}\ \bibnamefont
  {Chien}}, \bibinfo {author} {\bibfnamefont {D.}~\bibnamefont {Gruss}},
  \bibinfo {author} {\bibfnamefont {M.}~\bibnamefont {Di~Ventra}}, \ and\
  \bibinfo {author} {\bibfnamefont {M.}~\bibnamefont {Zwolak}},\ }\href@noop {}
  {\bibfield  {journal} {\bibinfo  {journal} {New J. Phys.}\ }\textbf {\bibinfo
  {volume} {15}},\ \bibinfo {pages} {063026} (\bibinfo {year}
  {2013})}\BibitemShut {NoStop}%
\bibitem [{\citenamefont {Chern}\ \emph {et~al.}()\citenamefont {Chern},
  \citenamefont {Chien},\ and\ \citenamefont {Di~Ventra}}]{GW14}%
  \BibitemOpen
  \bibfield  {author} {\bibinfo {author} {\bibfnamefont {G.~W.}\ \bibnamefont
  {Chern}}, \bibinfo {author} {\bibfnamefont {C.~C.}\ \bibnamefont {Chien}}, \
  and\ \bibinfo {author} {\bibfnamefont {M.}~\bibnamefont {Di~Ventra}},\
  }\href@noop {} {}\bibinfo {note} {Eprint, arXiv: 1307.6128}\BibitemShut
  {NoStop}%
\bibitem [{\citenamefont {Peotta}\ and\ \citenamefont
  {Di~Ventra}(2014)}]{Peotta2014}%
  \BibitemOpen
  \bibfield  {author} {\bibinfo {author} {\bibfnamefont {S.}~\bibnamefont
  {Peotta}}\ and\ \bibinfo {author} {\bibfnamefont {M.}~\bibnamefont
  {Di~Ventra}},\ }\href@noop {} {\bibfield  {journal} {\bibinfo  {journal}
  {Phys. Rev. A}\ }\textbf {\bibinfo {volume} {89}},\ \bibinfo {pages} {013621}
  (\bibinfo {year} {2014})}\BibitemShut {NoStop}%
\bibitem [{\citenamefont {Henderson}\ \emph {et~al.}(2009)\citenamefont
  {Henderson}, \citenamefont {Ryu}, \citenamefont {MacCormic},\ and\
  \citenamefont {Boshier}}]{ring2}%
  \BibitemOpen
  \bibfield  {author} {\bibinfo {author} {\bibfnamefont {K.}~\bibnamefont
  {Henderson}}, \bibinfo {author} {\bibfnamefont {C.}~\bibnamefont {Ryu}},
  \bibinfo {author} {\bibfnamefont {C.}~\bibnamefont {MacCormic}}, \ and\
  \bibinfo {author} {\bibfnamefont {M.~G.}\ \bibnamefont {Boshier}},\
  }\href@noop {} {\bibfield  {journal} {\bibinfo  {journal} {New J. Phys.}\
  }\textbf {\bibinfo {volume} {11}},\ \bibinfo {pages} {043030} (\bibinfo
  {year} {2009})}\BibitemShut {NoStop}%
\bibitem [{\citenamefont {Gaunt}\ \emph {et~al.}(2013)\citenamefont {Gaunt},
  \citenamefont {Schmidutz}, \citenamefont {Gotlibovych}, \citenamefont
  {Smith},\ and\ \citenamefont {Hadzibabic}}]{Boxpotential}%
  \BibitemOpen
  \bibfield  {author} {\bibinfo {author} {\bibfnamefont {A.~L.}\ \bibnamefont
  {Gaunt}}, \bibinfo {author} {\bibfnamefont {T.~F.}\ \bibnamefont
  {Schmidutz}}, \bibinfo {author} {\bibfnamefont {I.}~\bibnamefont
  {Gotlibovych}}, \bibinfo {author} {\bibfnamefont {R.~P.}\ \bibnamefont
  {Smith}}, \ and\ \bibinfo {author} {\bibfnamefont {Z.}~\bibnamefont
  {Hadzibabic}},\ }\href@noop {} {\bibfield  {journal} {\bibinfo  {journal}
  {Phys. Rev. Lett.}\ }\textbf {\bibinfo {volume} {110}},\ \bibinfo {pages}
  {200406} (\bibinfo {year} {2013})}\BibitemShut {NoStop}%
\bibitem [{\citenamefont {Susskind}\ and\ \citenamefont
  {Lindesay}(2005)}]{holography_book}%
  \BibitemOpen
  \bibfield  {author} {\bibinfo {author} {\bibfnamefont {L.}~\bibnamefont
  {Susskind}}\ and\ \bibinfo {author} {\bibfnamefont {J.}~\bibnamefont
  {Lindesay}},\ }\href@noop {} {\emph {\bibinfo {title} {The holographic
  universe: An introduction to black holes, information and the string theory
  revolution}}}\ (\bibinfo  {publisher} {World Scientific},\ \bibinfo {year}
  {2005})\BibitemShut {NoStop}%
\bibitem [{\citenamefont {Nielsen}\ and\ \citenamefont
  {Chuang}(2001)}]{QCbook}%
  \BibitemOpen
  \bibfield  {author} {\bibinfo {author} {\bibfnamefont {M.~A.}\ \bibnamefont
  {Nielsen}}\ and\ \bibinfo {author} {\bibfnamefont {I.~L.}\ \bibnamefont
  {Chuang}},\ }\href@noop {} {\emph {\bibinfo {title} {Quantum Computation and
  Quantum Information}}}\ (\bibinfo  {publisher} {Cambridge University Press},\
  \bibinfo {year} {2001})\BibitemShut {NoStop}%
\bibitem [{\citenamefont {Song}\ \emph {et~al.}(2011)\citenamefont {Song},
  \citenamefont {Flindt}, \citenamefont {Rachel}, \citenamefont {Klich},\ and\
  \citenamefont {Le~Hur}}]{KlichPRB11}%
  \BibitemOpen
  \bibfield  {author} {\bibinfo {author} {\bibfnamefont {H.~F.}\ \bibnamefont
  {Song}}, \bibinfo {author} {\bibfnamefont {C.}~\bibnamefont {Flindt}},
  \bibinfo {author} {\bibfnamefont {S.}~\bibnamefont {Rachel}}, \bibinfo
  {author} {\bibfnamefont {I.}~\bibnamefont {Klich}}, \ and\ \bibinfo {author}
  {\bibfnamefont {K.}~\bibnamefont {Le~Hur}},\ }\href@noop {} {\bibfield
  {journal} {\bibinfo  {journal} {Phys. Rev. B}\ }\textbf {\bibinfo {volume}
  {83}},\ \bibinfo {pages} {161408(R)} (\bibinfo {year} {2011})}\BibitemShut
  {NoStop}%
\bibitem [{\citenamefont {Song}\ \emph {et~al.}(2012)\citenamefont {Song},
  \citenamefont {Rachel}, \citenamefont {Flindt}, \citenamefont {Klich},
  \citenamefont {Laflorencie},\ and\ \citenamefont {Le~Hur}}]{KlichPRB12}%
  \BibitemOpen
  \bibfield  {author} {\bibinfo {author} {\bibfnamefont {H.~F.}\ \bibnamefont
  {Song}}, \bibinfo {author} {\bibfnamefont {S.}~\bibnamefont {Rachel}},
  \bibinfo {author} {\bibfnamefont {C.}~\bibnamefont {Flindt}}, \bibinfo
  {author} {\bibfnamefont {I.}~\bibnamefont {Klich}}, \bibinfo {author}
  {\bibfnamefont {N.}~\bibnamefont {Laflorencie}}, \ and\ \bibinfo {author}
  {\bibfnamefont {K.}~\bibnamefont {Le~Hur}},\ }\href@noop {} {\bibfield
  {journal} {\bibinfo  {journal} {Phys. Rev. B}\ }\textbf {\bibinfo {volume}
  {85}},\ \bibinfo {pages} {035409} (\bibinfo {year} {2012})}\BibitemShut
  {NoStop}%
\bibitem [{noQ()}]{noQSSC_note}%
  \BibitemOpen
  \href@noop {} {}\bibinfo {note} {For example, no QSSC is found in a tilted
  lattice potential. Moreover, if the initial filling has certain patterns (one
  particle per few sites), there is also no QSSC, as one can infer from
  Sec.~\ref{sec:MCFspatial}.}\BibitemShut {Stop}%
\bibitem [{\citenamefont {Zwolak}\ and\ \citenamefont
  {Di~Ventra}(2002)}]{DNAspintronics}%
  \BibitemOpen
  \bibfield  {author} {\bibinfo {author} {\bibfnamefont {M.}~\bibnamefont
  {Zwolak}}\ and\ \bibinfo {author} {\bibfnamefont {M.}~\bibnamefont
  {Di~Ventra}},\ }\href@noop {} {\bibfield  {journal} {\bibinfo  {journal}
  {App. Phys. Lett.}\ }\textbf {\bibinfo {volume} {81}},\ \bibinfo {pages}
  {925} (\bibinfo {year} {2002})}\BibitemShut {NoStop}%
\bibitem [{GEn()}]{GEnote}%
  \BibitemOpen
  \href@noop {} {}\bibinfo {note} {For the weak-link case, one may pick either
  the last (first) site on the left (right) lattice. The current is not
  affected by the choice if $E_{C}=E_{L,R}$ is chosen accordingly.}\BibitemShut
  {Stop}%
\bibitem [{\citenamefont {Heidrich-Meisner}\ \emph {et~al.}(2009)\citenamefont
  {Heidrich-Meisner}, \citenamefont {Manmana}, \citenamefont {Rigol},
  \citenamefont {Muramatsu}, \citenamefont {Feiguin},\ and\ \citenamefont
  {Dagotto}}]{Meisner09}%
  \BibitemOpen
  \bibfield  {author} {\bibinfo {author} {\bibfnamefont {F.}~\bibnamefont
  {Heidrich-Meisner}}, \bibinfo {author} {\bibfnamefont {S.~R.}\ \bibnamefont
  {Manmana}}, \bibinfo {author} {\bibfnamefont {M.}~\bibnamefont {Rigol}},
  \bibinfo {author} {\bibfnamefont {A.}~\bibnamefont {Muramatsu}}, \bibinfo
  {author} {\bibfnamefont {A.~E.}\ \bibnamefont {Feiguin}}, \ and\ \bibinfo
  {author} {\bibfnamefont {E.}~\bibnamefont {Dagotto}},\ }\href@noop {}
  {\bibfield  {journal} {\bibinfo  {journal} {Phys. Rev. A}\ }\textbf {\bibinfo
  {volume} {80}},\ \bibinfo {pages} {041603} (\bibinfo {year}
  {2009})}\BibitemShut {NoStop}%
\bibitem [{\citenamefont {Polkovnikov}\ \emph {et~al.}(2011)\citenamefont
  {Polkovnikov}, \citenamefont {Sengupta}, \citenamefont {Silva},\ and\
  \citenamefont {Vengalattore}}]{Polkovnikov11}%
  \BibitemOpen
  \bibfield  {author} {\bibinfo {author} {\bibfnamefont {A.}~\bibnamefont
  {Polkovnikov}}, \bibinfo {author} {\bibfnamefont {K.}~\bibnamefont
  {Sengupta}}, \bibinfo {author} {\bibfnamefont {A.}~\bibnamefont {Silva}}, \
  and\ \bibinfo {author} {\bibfnamefont {M.}~\bibnamefont {Vengalattore}},\
  }\href@noop {} {\bibfield  {journal} {\bibinfo  {journal} {Rev. Mod. Phys.}\
  }\textbf {\bibinfo {volume} {83}},\ \bibinfo {pages} {863} (\bibinfo {year}
  {2011})}\BibitemShut {NoStop}%
\bibitem [{\citenamefont {Beria}\ \emph {et~al.}(2013)\citenamefont {Beria},
  \citenamefont {Iqbal}, \citenamefont {Di~Ventra},\ and\ \citenamefont
  {Muller}}]{Beria13}%
  \BibitemOpen
  \bibfield  {author} {\bibinfo {author} {\bibfnamefont {M.}~\bibnamefont
  {Beria}}, \bibinfo {author} {\bibfnamefont {Y.}~\bibnamefont {Iqbal}},
  \bibinfo {author} {\bibfnamefont {M.}~\bibnamefont {Di~Ventra}}, \ and\
  \bibinfo {author} {\bibfnamefont {M.}~\bibnamefont {Muller}},\ }\href@noop {}
  {\bibfield  {journal} {\bibinfo  {journal} {Phys. Rev. A}\ }\textbf {\bibinfo
  {volume} {88}},\ \bibinfo {pages} {043611} (\bibinfo {year}
  {2013})}\BibitemShut {NoStop}%
\bibitem [{\citenamefont {Seaman}\ \emph {et~al.}(2007)\citenamefont {Seaman},
  \citenamefont {Kramer}, \citenamefont {Anderson},\ and\ \citenamefont
  {Holland}}]{Atomtronics07}%
  \BibitemOpen
  \bibfield  {author} {\bibinfo {author} {\bibfnamefont {B.~T.}\ \bibnamefont
  {Seaman}}, \bibinfo {author} {\bibfnamefont {M.}~\bibnamefont {Kramer}},
  \bibinfo {author} {\bibfnamefont {D.~Z.}\ \bibnamefont {Anderson}}, \ and\
  \bibinfo {author} {\bibfnamefont {M.~J.}\ \bibnamefont {Holland}},\
  }\href@noop {} {\bibfield  {journal} {\bibinfo  {journal} {Phys. Rev. A}\
  }\textbf {\bibinfo {volume} {75}},\ \bibinfo {pages} {023615} (\bibinfo
  {year} {2007})}\BibitemShut {NoStop}%
\bibitem [{\citenamefont {Sabetta}\ and\ \citenamefont
  {Misguich}(2013)}]{Sabetta13}%
  \BibitemOpen
  \bibfield  {author} {\bibinfo {author} {\bibfnamefont {T.}~\bibnamefont
  {Sabetta}}\ and\ \bibinfo {author} {\bibfnamefont {G.}~\bibnamefont
  {Misguich}},\ }\href@noop {} {\bibfield  {journal} {\bibinfo  {journal}
  {Phys. Rev. B}\ }\textbf {\bibinfo {volume} {88}},\ \bibinfo {pages} {245114}
  (\bibinfo {year} {2013})}\BibitemShut {NoStop}%
\bibitem [{\citenamefont {Antal}\ \emph {et~al.}(1999)\citenamefont {Antal},
  \citenamefont {Racz}, \citenamefont {Rakos},\ and\ \citenamefont
  {Schutz}}]{Antal99}%
  \BibitemOpen
  \bibfield  {author} {\bibinfo {author} {\bibfnamefont {T.}~\bibnamefont
  {Antal}}, \bibinfo {author} {\bibfnamefont {Z.}~\bibnamefont {Racz}},
  \bibinfo {author} {\bibfnamefont {A.}~\bibnamefont {Rakos}}, \ and\ \bibinfo
  {author} {\bibfnamefont {G.~M.}\ \bibnamefont {Schutz}},\ }\href@noop {}
  {\bibfield  {journal} {\bibinfo  {journal} {Phys. Rev. E}\ }\textbf {\bibinfo
  {volume} {59}},\ \bibinfo {pages} {4912} (\bibinfo {year}
  {1999})}\BibitemShut {NoStop}%
\bibitem [{\citenamefont {Thomas}\ and\ \citenamefont
  {Flindt}(2014)}]{Waitingtime14}%
  \BibitemOpen
  \bibfield  {author} {\bibinfo {author} {\bibfnamefont {K.~H.}\ \bibnamefont
  {Thomas}}\ and\ \bibinfo {author} {\bibfnamefont {C.}~\bibnamefont
  {Flindt}},\ }\href@noop {} {\bibfield  {journal} {\bibinfo  {journal} {Phys.
  Rev. B}\ }\textbf {\bibinfo {volume} {89}},\ \bibinfo {pages} {245420}
  (\bibinfo {year} {2014})}\BibitemShut {NoStop}%
\bibitem [{pop()}]{population_note}%
  \BibitemOpen
  \href@noop {} {}\bibinfo {note} {For smaller initial bias, the distributions
  on the two sides still evolve dynamically and a quasi steady-state current is
  maintained for a certain period, but an inversion of the population on the
  two sides may not occur during that period.}\BibitemShut {Stop}%
\bibitem [{\citenamefont {Cheneau}\ \emph {et~al.}(2012)\citenamefont
  {Cheneau}, \citenamefont {Barmettler}, \citenamefont {Poletti}, \citenamefont
  {Endres}, \citenamefont {Schaub}, \citenamefont {Fukuhara}, \citenamefont
  {Gross}, \citenamefont {Bloch}, \citenamefont {Kollath},\ and\ \citenamefont
  {Kuhr}}]{LC1}%
  \BibitemOpen
  \bibfield  {author} {\bibinfo {author} {\bibfnamefont {M.}~\bibnamefont
  {Cheneau}}, \bibinfo {author} {\bibfnamefont {P.}~\bibnamefont {Barmettler}},
  \bibinfo {author} {\bibfnamefont {D.}~\bibnamefont {Poletti}}, \bibinfo
  {author} {\bibfnamefont {M.}~\bibnamefont {Endres}}, \bibinfo {author}
  {\bibfnamefont {P.}~\bibnamefont {Schaub}}, \bibinfo {author} {\bibfnamefont
  {T.}~\bibnamefont {Fukuhara}}, \bibinfo {author} {\bibfnamefont
  {C.}~\bibnamefont {Gross}}, \bibinfo {author} {\bibfnamefont
  {I.}~\bibnamefont {Bloch}}, \bibinfo {author} {\bibfnamefont
  {C.}~\bibnamefont {Kollath}}, \ and\ \bibinfo {author} {\bibfnamefont
  {S.}~\bibnamefont {Kuhr}},\ }\href@noop {} {\bibfield  {journal} {\bibinfo
  {journal} {Nature}\ }\textbf {\bibinfo {volume} {481}},\ \bibinfo {pages}
  {484} (\bibinfo {year} {2012})}\BibitemShut {NoStop}%
\bibitem [{\citenamefont {Geiger}\ \emph {et~al.}(2014)\citenamefont {Geiger},
  \citenamefont {Langen}, \citenamefont {Mazets},\ and\ \citenamefont
  {Schmiedmayer}}]{Geiger14}%
  \BibitemOpen
  \bibfield  {author} {\bibinfo {author} {\bibfnamefont {R.}~\bibnamefont
  {Geiger}}, \bibinfo {author} {\bibfnamefont {T.}~\bibnamefont {Langen}},
  \bibinfo {author} {\bibfnamefont {I.}~\bibnamefont {Mazets}}, \ and\ \bibinfo
  {author} {\bibfnamefont {J.}~\bibnamefont {Schmiedmayer}},\ }\href@noop {}
  {\bibfield  {journal} {\bibinfo  {journal} {New J. Phys.}\ }\textbf {\bibinfo
  {volume} {16}},\ \bibinfo {pages} {053034} (\bibinfo {year}
  {2014})}\BibitemShut {NoStop}%
\bibitem [{\citenamefont {Kubo}\ \emph {et~al.}(2004)\citenamefont {Kubo},
  \citenamefont {Toda},\ and\ \citenamefont {Hashitsume}}]{Kubo_book}%
  \BibitemOpen
  \bibfield  {author} {\bibinfo {author} {\bibfnamefont {R.}~\bibnamefont
  {Kubo}}, \bibinfo {author} {\bibfnamefont {M.}~\bibnamefont {Toda}}, \ and\
  \bibinfo {author} {\bibfnamefont {N.}~\bibnamefont {Hashitsume}},\
  }\href@noop {} {\emph {\bibinfo {title} {Statistical Physics II:
  Nonequilibrium Statistical Mechanics}}},\ \bibinfo {edition} {2nd}\ ed.\
  (\bibinfo  {publisher} {Springer-Verlag},\ \bibinfo {address} {Berlin},\
  \bibinfo {year} {2004})\BibitemShut {NoStop}%
\bibitem [{\citenamefont {Fetter}\ and\ \citenamefont
  {Walecka}(2003)}]{Fetter_book}%
  \BibitemOpen
  \bibfield  {author} {\bibinfo {author} {\bibfnamefont {A.~L.}\ \bibnamefont
  {Fetter}}\ and\ \bibinfo {author} {\bibfnamefont {J.~D.}\ \bibnamefont
  {Walecka}},\ }\href@noop {} {\emph {\bibinfo {title} {Quantum Theory of
  Many-Particle Systems}}}\ (\bibinfo  {publisher} {Dover Publications},\
  \bibinfo {address} {New York},\ \bibinfo {year} {2003})\BibitemShut {NoStop}%
\bibitem [{\citenamefont {Jauho}\ \emph {et~al.}(1994)\citenamefont {Jauho},
  \citenamefont {Wingreen},\ and\ \citenamefont {Meir}}]{Jauho94}%
  \BibitemOpen
  \bibfield  {author} {\bibinfo {author} {\bibfnamefont {A.~P.}\ \bibnamefont
  {Jauho}}, \bibinfo {author} {\bibfnamefont {N.~S.}\ \bibnamefont {Wingreen}},
  \ and\ \bibinfo {author} {\bibfnamefont {Y.}~\bibnamefont {Meir}},\
  }\href@noop {} {\bibfield  {journal} {\bibinfo  {journal} {Phys. Rev. B}\
  }\textbf {\bibinfo {volume} {50}},\ \bibinfo {pages} {5528} (\bibinfo {year}
  {1994})}\BibitemShut {NoStop}%
\end{thebibliography}
%

\end{document}